\documentclass[a4paper,11pt]{article}
\pdfoutput=1 % if your are submitting a pdflatex (i.e. if you have 
\usepackage{jheppub} % for details on the use of the package, please
            
\usepackage{bm,amssymb,slashed,graphicx,multirow,soul,mathtools,xspace,array,comment,extarrows}  
\usepackage{float}                   
\allowdisplaybreaks 
\usepackage{ bbold }
\usepackage{color}
\usepackage{subcaption}
\usepackage[printonlyused]{acronym}
\usepackage{hyperref}
\usepackage{overpic}
\usepackage{hhline}
\acrodef{PDG}[PDG]{Particle Data Group}
\acrodef{OPE}[OPE]{Operator Product Expansion}
\acrodef{FCNC}[FCNC]{flavour-changing neutral current}
\acrodef{RHC}[RHC]{right-handed currents}
\acrodef{SM}[SM]{Standard Model}
\acrodef{NP}[NP]{New Physics}
\acrodef{MFV}[MFV]{Minimal Flavour Violation}
\acrodef{SD}[SD]{short-distance}
\acrodef{LD}[LD]{long-distance}
\acrodef{DA}[DA]{distribution amplitude}

\newcommand{\PHOTOS}{\texttt{PHOTOS}\xspace}
\newcommand{\PHOTONS}{\texttt{PHOTONS++}\xspace}
\newcommand{\SHERPA}{\texttt{SHERPA}\xspace}

\newcommand{\FB}{F_B}

\newcommand{\pperp}[1]{P^\perp_{#1}}
\newcommand{\ppara}[1]{P^\parallel_{#1}}
\newcommand{\pB}[1]{(p_B)_{#1}}

\newcommand{\VEV}[1]{\langle #1 \rangle}

\newcommand{\decut}{E_\ga^{\text{cut}}}

\newcommand{\matel}[3]{\langle #1|#2|#3\rangle}
\newcommand{\al}{\alpha}
\newcommand{\be}{\beta}
\newcommand{\ga}{\gamma}
\newcommand{\de}{\delta}

\newcommand{\eps}{\epsilon}

\newcommand{\GeV}{\,\mbox{GeV}}
\newcommand{\MeV}{\,\mbox{MeV}}
\newcommand{\ORD}{{\cal O}}

\newcommand{\TAB}{Tab.~}
\newcommand{\FIG}{Fig.~}
\newcommand{\FIGs}{Figs.~}
\newcommand{\SEC}{Sec.~}

\newcommand{\APP}{App.~}

\newcommand{\EQ}{Eq.~}

\newcommand{\Tr}{\mathrm{Tr}}

\newcommand{\KallenpB}{ \lambda_{p_B}}
\newcommand{\Kallenq}{ \lambda_{q}}

\newcommand{\para}{\parallel}

\newcommand{\Amp}{{\cal A}}

\newcommand{\epsIR} {\eps_{\textrm{IR}}}

\newcommand{\geff}{g_{\textrm{eff}}}

\newcommand{\hc}{{\text{hc}}}

\newcommand{\phiB}{{\Phi_{B}}}

\newcommand{\phiBpl}{{\Phi_{B^-}}}
\newcommand{\momB}{{p}_{\Phi}}

\newcommand{\rr}{r}

\newcommand{\lepB}{B^- \to \ell^- \bar \nu}

\newcommand{\Hpz}{{\cal J}^{} _B}
\newcommand{\Hpzda}{{\cal J}^{\dagger}_B}

\newcommand{\ZBp}{{\cal Z}_B^{}}

\usepackage{datatool}

\newcommand{\lscale}[1]{{ |#1\! \GeV}}

\newcommand{\gphi}{\ZBp}

\newcommand{\qbarq}{\langle \bar q q \rangle}

\newcommand{\gtilde}{\tilde{g}}

\usepackage{physics}

\definecolor{violet}{rgb}{0.94, 0.2, 0.8}%lightcoral
\definecolor{lightblue}{rgb}{0.39, 0.58, 1.00} %cornflowerblue
\definecolor{lightgreen}{rgb}{0.1, 0.73, 0.33}

\newcommand{\VA}{V-A }
\newcommand{\SP}{S-P }
\newcommand{\bl}{b\,\ell_1 }

\numberwithin{equation}{section}

\title{\boldmath Structure-dependent QED in $B^- \to \ell^- \bar \nu (\ga)$}
\author[1]{Matthew Rowe,}\author[1]{Roman Zwicky}

 \affiliation[1]{Higgs Centre for Theoretical Physics, School of Physics and Astronomy, University of Edinburgh, Edinburgh, EH9 3JZ, Scotland}

\emailAdd{m.j.rowe@sms.ed.ac.uk}
\emailAdd{roman.zwicky@ed.ac.uk}

\abstract{
Based on  explicitly gauge invariant interpolating operators we compute complete next-leading order QED-corrections for 
leptonic decays. These are sizeable since the helicity-suppression in V-A interactions allows for structure-dependent 
collinear logs. We have explicitly checked that these logs are absent for helicity-unsuppressed Yukawa-type transitions. 
Based on $B \to \gamma$ form factors we present the rates for $B^- \to (\mu^-,\tau^-) \bar \nu (\gamma)$ in differential and integrated form as a function of the photon energy cut-off $\decut$.
 The effect of the virtual structure-dependent 
corrections are approximately   $+5\%$ and $+3\%$  for the $\mu$- and $\tau$-channel respectively. The structure dependence of the real radiation exceeds that of the virtual one for $\decut|_{\mu}  > 0.18(3)$ GeV and is subdominant for the tau channel even when fully inclusive.
}

\begin{document}

%\toccontinuoustrue

\maketitle

\newpage

\flushbottom

\setcounter{tocdepth}{3}
\setcounter{page}{1}
\pagestyle{plain}

%\pagenumbering{arabic}

\section{Introduction}
\label{sec:intro}

It has been known for a long time that QED challenges the notion of a charged  elementary particle 
because of the emission of infinitely many soft photons leading to seemingly infrared (IR) divergent processes. 
 At the conceptual level this problem has been resolved  by abandoning the one-particle 
Fock space  in favour of  the coherent state formalism \cite{Chung:1965zza,Kibble:1968oug,Kibble:1968npb,Kibble:1968lka,Kulish:1970ut}.  A more pragmatic solution 
is the well-known Bloch-Nordsieck cancellation mechanism \cite{Bloch:1937pw}, 
whereby virtual corrections and real emissions 
processes are added consistently at each order in the fine structure constant $\al \equiv e^2/(4 \pi) \approx 1/137$ giving rise  to IR-finite processes.  
As experiment is seldom fully inclusive  the photon-energy is either cut-off  explicitly or one considers the 
photon-energy differential rate.   This leads to large soft-logs but since 
they originate from the  soft region the structure of the particles is not resolved and thus treating the latter 
in  the point-like approximation (i.e. scalar QED) suffices  and is well-understood 
including resummation  \cite{Yennie:1961ad,Weinberg:1965nx}. The other sizeable logs 
are the collinear ones which arise when the photon is collinear to a light charged particle, 
 such as an electron in a $B$-decay. For example, in $B \to K e^+ e^-$ collinear logs of the type 
 $\frac{\al}{\pi} \ln (m_e/m_B)$ lead to corrections 
of up to $10$-$20\%$ depending on the phase space region \cite{Isidori:2020acz,Isidori:2022bzw}. 
 Since the collinearity does not restrain the photon energy the point-like approximation is not applicable 
 and  structure-dependence can give rise to large (hard)-collinear logs.  
 However, using gauge invariance and the KLN-theorem \cite{Kinoshita:1962ur,Lee:1964is}, which extends the Bloch-Nordsieck  idea to massless charged particles, it was shown that no such logs can arise \cite{Isidori:2020acz}. There are though exceptions: i) when the collinear charged particle is composite e.g $\pi^-$ \emph{or} 
 ii) when the KLN-theorem does not need to hold.  
 It is precisely the second exception that applies to $B^- \to \ell^-\bar \nu$-type decays mediated by 
 \VA interaction as in the Standard Model (SM).  The point is that  the amplitude is chirally suppressed 
  $\Amp \propto m_\ell \ln m_\ell$ and thus  finite for $m_\ell \to 0$ and therefore blind to 
 the KLN-theorem which is based on unitarity cancelling IR divergences  \cite{Zwicky:2021olr,Nabeebaccus:2022jhu}.  
 
 To resolve hadrons one then needs a non-perturbative formalism. Possibly the most natural setting 
 is chiral perturbation theory (ChPT)  
 as it  extends the point-like approximations through its  energy expansion (e.g. 
   \cite{Knecht:1999ag,Cirigliano:2007ga,Gasser:2010wz,Cirigliano:2011tm}  
 and  \cite{Cetal01,CGH08,Descotes-Genon:2005wrq}  for leptonic and semi-leptonic decays respectively). 
 However it does not apply 
 to heavy decays and the  other methods such as lattice QCD  face, one way or another, the challenge that  
 the standard interpolating operator  for a $B_q$-meson 
 \begin{equation}
 \label{eq:naive}
J_B = m_+\,  \bar b i \gamma_5 q \; , \qquad m_{\pm} = m_b \pm m_q \;, \quad \matel{ B^-}{J_B}{0} = m_B^2 f_B \; ,
\end{equation}
is not QED gauge invariant for a charged $B_q$-meson.
In the context of QCD sum rules  a solution was proposed 
\cite{Nabeebaccus:2022jhu} by introducing a long distance (on-shell) $B$-meson in terms of  a scalar field $\phiB$ 
for which the generalised LSZ-factor has been shown to be IR-finite. 
In this paper we explicitly verify that all the universal collinear and soft logs  are reproduced separately for the
 virtual and  the real rates thereby validating the approach.   
 On the lattice the interpolating operator problem  disappears in the limit of infinite Euclidean time  separation of 
the meson source as the exact LSZ formula emerges 
\cite{Carrasco:2015xwa,Endres:2015gda,Feng:2018qpx}.\footnote{In addition there is 
an explicit gauge invariant formulation using so-called $C^*$-boundary conditions \cite{Lucini:2015hfa}.} 
Numerical QED studies include virtual effects in $(\pi,K) \to \ell \bar \nu$ \cite{Lubicz:2016mpj}, including real 
radiation  \cite{DiCarlo:2019thl}, real radiation in $D_s \to \ell \bar  \nu $ decays  \cite{Desiderio:2020oej,Frezzotti:2023ygt}  and now very recently for $B_s \to \mu^+\mu^- (\ga)$ \cite{Frezzotti:2024kqk}.
QED effects have also been investigated in 
soft-collinear effective theory (SCET)  \cite{BBS17,Beneke:2019slt} and specifically for  $B^- \to \ell^-\bar \nu$ \cite{Cornella:2022ubo} where the 
interpolating operator problem is replaced by defining process-dependent light-cone distribution amplitudes   \cite{Beneke:2021pkl,Beneke:2022msp}.
 
In experiment QED-corrections are dealt with by using Monte-Carlo  generators such as \PHOTOS  
\cite{Barberio:1990ms,Barberio:1993qi,Golonka:2005pn,PHOTOS}, or the \PHOTONS  module \cite{Schonherr:2008av} of \SHERPA \cite{Sherpa:2019gpd} for which $B^- \to \ell^-\bar \nu$  structure-dependent corrections have not yet been implemented since they are unknown.\footnote{Hence current Monte-Carldo 
QED-effects are based on scalar QED (point-like approximation).} In this paper we close this gap by 
computing the virtual structure-dependent corrections for $B^- \to \ell^-\bar \nu$, applying the above mentioned 
framework  \cite{Nabeebaccus:2022jhu}.
The structure-dependent part of the real radiation is described by the two $B\to \ga$ form factors computed in sum rules on the light-cone at NLO in 
$\al_s$ \cite{Janowski:2021yvz}.\footnote{Whereas the standard application of these form factors are to 
describe the phenomenology of  $\bar{B}^- \to \ell^- \bar \nu \ga$ \cite{Korchemsky:1999qb,Lunghi:2002ju,DescotesGenon:2002mw,Bosch:2003fc,Beneke:2011nf,Braun:2012kp,Wang:2016qii,Beneke:2018wjp,Shen:2018abs,Wang:2018wfj} 
or $B_{d,s} \to \ell^+\ell^- \ga$  \cite{Dettori:2016zff,Aditya:2012im,GRZ17,Kozachuk:2017mdk,Beneke:2020fot,Carvunis:2021jga,MN04,Kozachuk:2017mdk,Beneke:2020fot} decays, they equally serve as real radiation in the soft region for $\bar{B}^- \to \ell^- \bar \nu$ and  $B_{d,s} \to \ell^+\ell^-$ 
respectively.} 
Including structure-dependent virtual and real contributions from different sources is not a problem since they are separately IR-finite (unlike scalar QED). The numerical effects relative to scalar QED, which is sizeable by itself, are
found to be relevant with respect to other inputs such as the   
$B$-meson decay constant  and $|V_{\rm ub}|$ (cf.  \SEC\ref{sec:outlook}).
While we provide corrections for the $B^- \to \ell^-\bar \nu (\ga)$, our framework extends to $D$- and possibly $K$-mesons. 
Neutral decays modes such as $B_s \to \mu^+\mu^-$ are  simpler as then the long-distance field $\phiB$ is not required but
the kinematics require a slightly different computation.
 % to which we may return to in a future 
%publication. 

The paper is organised as follows. In \SEC\ref{sec:phiB} the conceptual framework is summarised, 
followed by the computation in 
\SEC\ref{sec:Comp} and the discussion of the collinear logs in \SEC\ref{sec:Coll}.
Numerical results including comparison to the literature are presented in \SEC\ref{sec:num}. 
The paper ends with conclusions and a summary in \SEC\ref{sec:con}.  
Appendices contain for example the computational results of the condensates, 
a discussion on the necessity of all cuts and further plots.

\section{The gauge invariant formalism for $B^- \rightarrow \ell^- \bar \nu (\gamma)$}
\label{sec:phiB}

As stated in the introduction the  key change is the introduction of a long-distance $B$-meson $\phiB$ to 
render the interpolating current \eqref{eq:naive} gauge invariant  \cite{Nabeebaccus:2022jhu}
\begin{equation}
\label{eq:JBp} 
 \Hpz  \equiv J_B  \phiB \;, \quad 
\ZBp  \equiv \matel{B^- }{\Hpz}{\phiBpl }  \;.
\end{equation}
We drop  the ``$(0)$" superscript with respect to the above-mentioned  reference.  It is useful to define 
 the charge assignments to the currents
\begin{equation}
\label{eq:Q}
Q_{\Hpz} \equiv Q_\Phi + Q_q - Q_b \;, \qquad  Q_{J_B} \equiv   Q_q - Q_b \;.
\end{equation}
The matrix element $\ZBp$   takes on  the r\^ole of the LSZ-factor and has been shown to be IR-finite  
and (explicitly) gauge invariant \cite{Nabeebaccus:2022jhu}. Its  key elements will be discussed further
below.
The decay rate, with the Bloch-Nordsieck cancellation mechanism in place is given by
 \begin{eqnarray}
 \label{eq:master}
& &  \Gamma (\lepB (\ga)) =  \nonumber \\[0.2cm]
& &  \qquad  \frac{1}{   |\ZBp|^2  } \times
 \int_{\decut} d \Phi_\ga  | \ZBp|^2 \left(  | {\cal A}(\lepB ) |^2 \de(\Phi_\ga)   +|  {\cal A}(\lepB \ga) |^2 \right) \;,
 \end{eqnarray}
 where  $\decut  > E_\ga $ is the photon energy cut-off, in the rest frame of the decaying 
 particle.
 The delta function $ \de(\Phi_\ga) $   in the photon kinematic variables, to be made more precise further below,
 singles out the virtual contribution.
 The point of the sum rule procedure is that both the LSZ-factor $|\ZBp|^2$ and the integrand 
 are computed separately from appropriate correlation functions as outlined  below.

\subsection{The main process}
\label{sec:main}

The integrand    in \eqref{eq:master},  
can be extracted from the following correlation function
\begin{eqnarray}
\label{eq:mainSR}
\Pi^{(\ga)}(p_B^2, \momB^2) &\;=\;& i  \int _x e^{i x \rr}  \matel{ \ell \bar \nu (\ga) }{ T 
   {\Hpz}   (x) (- i {\cal L}_W(0) ) 
 }{  \phiB( \momB)}   \nonumber \\[0.1cm]
&\;=\;&  \int \frac{ds}{2\pi i} \frac{ \text{disc}_s[\Pi^{(\ga)}(s, \momB^2) ]}{s-p_B^2-i0} =   \frac{ \ZBp \, i  {\cal A}( 
\lepB (\ga))    }{ m_B^2-p_B^2} + \dots \;,
 \end{eqnarray}
which we have written as a dispersion relation of which the residuum of the lowest lying state carries 
 the information of interest. The  dots denote  higher states in the spectrum which will be suppressed 
 by the usual Borel transform.
Above $\int_x \equiv \int \dd[d]{x}$ is a shorthand used hereafter, 
$\mathcal{A}^{(\gamma)} \equiv \matel{\ell \bar \nu (\gamma)}{-\mathcal{L}_W}{B^-}$ the amplitude 
and ${\cal L}_W$   the electroweak Lagrangian
 \begin{equation}
\label{eq:weakH}
    \mathcal{L}_W = - g_{\text{eff}} \, \bar \ell \Gamma^\mu \nu \, \bar u \hat{\Gamma}_\mu b\;, \qquad \Gamma_\mu = \gamma_\mu (C_V + C_A \gamma_5) \;, \qquad \hat{\Gamma}_\mu = \gamma_\mu (1-\gamma_5) \;,
\end{equation}
with $C_{V(A)} = 1(-1)$ in the SM 
\begin{equation}
\label{eq:mom}
   p_\Phi  =    p_B + r \;,  \qquad q = p_B - k \;, \qquad  \geff = - \frac{G_F}{\sqrt{2}}V_{\text{ub}} \;,
\end{equation}
with $G_F$ the Fermi constant and $V_{\text{ub}}$ the CKM matrix element governing the $b \to u$ transition. 
The momenta $r$ is auxiliary as it is used to distinguish between the on-shell $p_\Phi$ and the off-shell $p_B$ momenta 
used for the dispersion relation. {We comment on its role and how it finally does not play a role further below.}
The $b \bar u$-flavour is chosen for concreteness and replacement by other flavour structures is 
straightforward.  {It is important to distinguish $p_\Phi$  in the real and virtual case}
\begin{equation}
 p_\Phi|_{\text{virt}} = l_1 + l_2 \;,  \quad p_\Phi|_{\text{real }} = l_1 + l_2 + k \;,
\end{equation}
where $l_{1,2}$ and $k$ are lepton and photon momenta respectively. 
% Before becoming more concrete about the computation  we want to get across 
%the meaning and treatment of $\phiB$ and why it is interesting to consider interactions other than the V-A. 

\paragraph{Theoretical considerations:} If the meson that $J_B$ interpolates was stable and we were able to apply a perfect LSZ projection 
then there would be no need to be concerned with gauge dependence as it would  simply  cancel 
in the ratio \eqref{eq:master}. In other words all artefacts are contained in $|\ZBp|$. The problem is that a perfect LSZ can never be achieved in practice for a composite particle in a non-trivial theory and thus the necessity to act. Concretely, in QCD sum rules $J_B$ is not perfectly on-shell since the dispersive integral 
is approximated by a perturbative computation. 
This is where $\phiB( p_\Phi)$ comes in as it is on-shell and the auxiliary momenta 
$r$ in \eqref{eq:mom}  parameterises that. In order to recover the correct soft logs one must ensure   \cite{Nabeebaccus:2022jhu}
\begin{equation}
\label{eq:onshell}
p_\Phi^2 = m_B^2 \;,
\end{equation}
  which intuitively seems the most natural choice in any case.
%The goal of the computation is to make $r$ disappear in the end. 
%How this is achieved is detailed further below. 
 
Let us turn to why it is interesting to consider, besides the \VA interaction, the \SP interaction of Yukawa-type. 
As previously stated, the reason there are structure-dependent logs in the \VA case is that the amplitude 
is chirally suppressed. This suppression is lifted in the \SP case. Symbolically we have
\begin{alignat}{3}
& \Amp( B \to  \ell \bar \nu)|_{\text{V-A}} &\;\propto\;& \ORD(m_\ell) &\;+\;&  \ORD( \al)   m_\ell \ln m_\ell \;,\nonumber \\[0.1cm]
& \Amp( B \to  \ell \bar \nu)|_{\text{S-P}} &\;\propto\;& \ORD(m^0_\ell ) &\;+\;&  \ORD( \al) m^0_\ell   \ln m_\ell \;,
\end{alignat}
abbreviating $ B^- \to  \ell^- \bar \nu  $ to $ B \to \ell  \bar  \nu$ here and occasionally elsewhere.
Hence the \SP can serve as a test case to reproduce the hard-collinear logs as done and advertised in 
\cite{Zwicky:2021olr}. The hard-collinear logs can 
be reproduced using the splitting function.  Formally the \SP case simply  follows through  substitution 
$\ga_\mu \to \mathbb{1}$ for  $\hat{\Gamma}_\mu$ and $ \Gamma_\mu$  in \eqref{eq:weakH}. 
Note that for the real radiation 
the helicity-suppression is lifted for the V-A interactions.  
Hence, for the SM case there are then two effects due to  the helicity-suppression: 
\begin{itemize}
\item [i)] large hard-collinear logs for the virtual  structure-dependent QED-corrections 
\item [ii)] significant enhancement of the real radiation for large $\decut$ 
\end{itemize}
Where ii) is well-appreciated point i) is not and its clarification is one of the main points of this paper. 
These effects are of course more pronounced for the muon since the tau mass is rather large.

\begin{centering}
\begin{figure}
\begin{overpic}[width=1.0\linewidth]{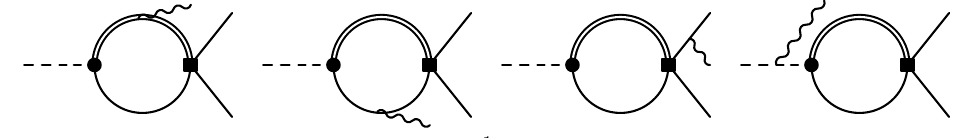}  
\put(4,8.5){$\phiB$}
\put(10,12){$b$}
\put(14.5,0.5){$\bar u$}
\put(25,14){$\ell^-$}
\put(25,0.5){$\bar \nu$}
\put(20,14){$\gamma$}
\end{overpic}
	\caption{\small  Diagrams contributing to $\Pi^{\ga}(p_B^2,\momB^2)$ in \eqref{eq:mainSR} 
	(i.e. the radiative or real emission part). The last diagram is specific to the $\phiB$-particle. The black circle denotes the $\Hpz$ operator, while the black square denotes the weak vertex $\mathcal{L}_w$. We stress that for the real structure dependent part we take external form factors \cite{Janowski:2021yvz} which 
	include many more contributions to the above diagrams. 
	E.g. higher twist in the photon distribution amplitude and  $\mathcal{O}(\alpha_s)$.}
	\label{fig:dia-real-nom}
\end{figure}
\end{centering}

\begin{centering}
\begin{figure}[h!]
\begin{overpic}[width=1.0\linewidth]{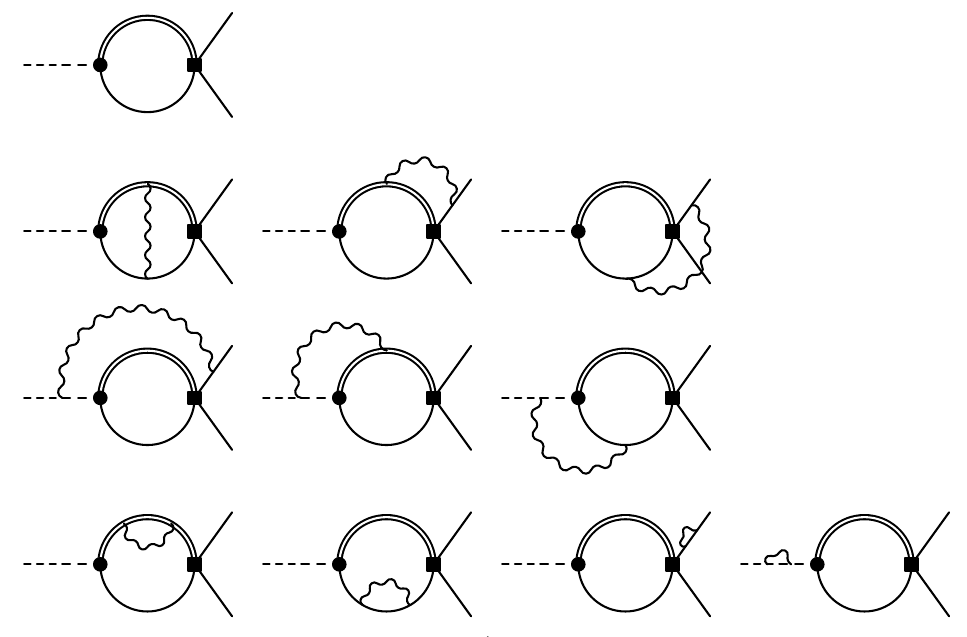}  
\put(4,60){$\phiB$}
\put(14.5,65.5){$b$}
\put(14.5,52){$\bar u$}
\put(25,65){$\ell^-$}
\put(25,52){$\bar \nu$}
\put(2,48){$b\,q$}
\put(27,48){$b\,\ell_1$}
\put(52,48){$q\,\ell_1$}
\put(2,33){$\Phi \, \ell_1$}
\put(27,33){$\Phi \, b$}
\put(52,33){$\Phi \, q$}
\put(2,13){$b \, b$}
\put(27,13){$q \, q$}
\put(52,13){$\ell_1 \, \ell_1$}
\put(77,13){$\Phi \, \Phi$}
\end{overpic}
	\caption{\small  Diagrams contributing to $\Pi(p_B^2,\momB^2)$ in \eqref{eq:mainSR} 
	(ie. the non-radiative part, hence no $(\ga)$ superscript). Top line is the LO diagram and the third line and the last diagram are specific to the $\phiB$-particle.}
	\label{fig:dia-virt-nom}
\end{figure}
\end{centering}

\paragraph{The computation:}  
The amplitude times the LSZ-factor, the square root of the integrand, 
 is then obtained by the usual sum rule procedure
\begin{equation} 
\label{eq:Areal}
 \, i {\cal A}( \lepB (\ga))    = 
 \frac{1}{ \ZBp} \int_{m_+^2}^{s_0} ds  e^{\frac{(m_B^2-s)}{M^2}} \rho_{\Pi^{(\ga)}}(s)  \;,
\end{equation}
where $2 \pi  i \rho_\Pi(s) = 2i \Im \Pi (s,m_B^2) =  \text{disc}_s \Pi(s, m_B^2   )$ by analyticity (and the Schwartz reflection principle). In the above $M^2$ and $s_0$ are the Borel parameter and continuum threshold which are the sum rule 
 specific parameters.  It is convenient to decompose this 
 quantity  into  LO and NLO in $\al$ parts and  perturbative versus quark condensate parts
\begin{equation}
\label{eq:below}
\rho_{\Pi^{(\ga)}} (s) = \rho^{(0)}_{{\Pi^{(\ga)}} ,\mathbb{1}}(s) + 
 \VEV{\bar qq} \rho^{(0)}_{{\Pi^{(\ga)}} ,\bar qq}(s)  + \frac{\al}{\pi}  ( \rho^{(2)}_{{\Pi^{(\ga)}} ,\mathbb{1}}(s)+ 
 \VEV{\bar qq} \rho^{(2)}_{{\Pi^{(\ga)}} ,\bar qq}(s)) \;.
\end{equation}
The quantity $\rho_{\Pi}$ is an implicit function of $\{ m_B^2, m_b, m_q, m_\ell \}$,
as well as the photon energy $E_\ga$ and the photon angle $c_\ga$ (to be made precise later) 
in the case of 
the real emission.  The real  emission and virtual diagrams are depicted 
in \FIG\ref{fig:dia-real-nom} and \FIG\ref{fig:dia-virt-nom} respectively. {It is observed that due to the presence 
of the (charged) $\phiB$ the number of diagrams  proliferates.} 
As discussed in \SEC \ref{sec:realComp} the calculation for the real radiative rate can essentially be recast in terms of $B \rightarrow \gamma$ form factors which contain all the structure-dependence.   
{We take the light-cone sum rule (LCSR) computation in  \cite{Janowski:2021yvz}  which is  NLO in $\al_s$ and itself valid for $q^2 < 14 \GeV^2$. This computation has been extended to 
the $B^*(B_1)$-pole, using the residue from a  different (but related computation)  \cite{Pullin:2021ebn}, 
 with a  $z$-parametrisaton ansatz  \cite{Janowski:2021yvz}.   
In fact it is the proximity to the pole of our domain of interest which makes it hard to impossible 
to compute the form factor directly with perturbative methods.}

\begin{centering}
\begin{figure}[h!]
\begin{overpic}[width=1.0\linewidth]{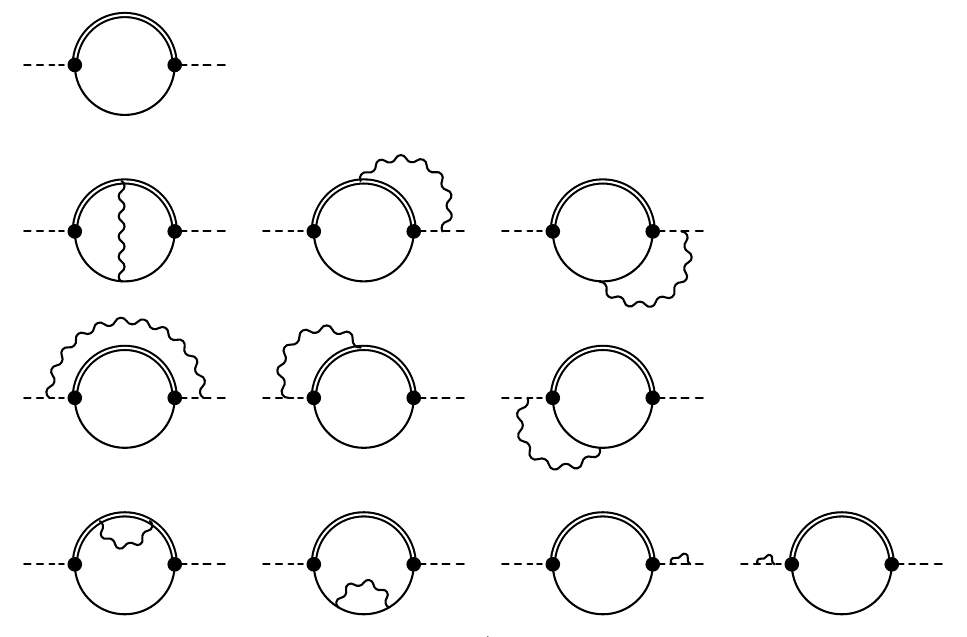}  
\put(2,60){$\phiB$}
\put(12,65.5){$b$}
\put(12,52){$\bar u$}
\put(20.5,60){$\phiB$}
\end{overpic}
	\caption{\small  Diagrams contributing to $C(p_B^2, \momB^2)$ \eqref{eq:C}, 
	that is $|\ZBp|^2$ \eqref{eq:cZB}. Again the top line is the LO contribution.}
	\label{fig:dia-denom}
\end{figure}
\end{centering}
\paragraph{On the auxiliary momenta $r$:}   How does the unphysical momenta  $r$ disappear from the final calculation? First, 
 Lorentz structures involving $r$ that appear in the calculation such as $\bar u \slashed{r} v$, $r \cdot \epsilon^*$ etc are considered extra unphysical information and can be discarded. 
 Second, scalar products involving $r$ are $\ORD(p_B^2 - m_B^2)$ 
 and would vanish in the case of a perfect LSZ. In the sum rule  they give rise to  small effects since the dispersion integral enforces $p_B^2 \approx m_B^2$ numerically 
 and they will enter the  uncertainty budget. 
 Third,  the  freedom in choosing other scalar products such as $r^2$ and $l_2 \cdot r $  allows us to set
 \begin{equation}
 r^2 =0 \;, \quad l_2 \cdot r = 0 \;.
 \end{equation}
 {The choice $l_2 \cdot r = 0$, simply  serves to remove the auxiliary momenta $r$ and 
 is in particular feasible since the neutrino is massless (the scalar product of two 
 lightlike vectors can be set to zero). 
 All of these concepts were  tested in a toy model  presented in \APP\ref{app:toy}. 
 The reader might find it helpful to disentangle the inner workings of this approach within 
 that simplified setting. We stress that in both the toy model and our hadronic framework all 
  universal IR logs have been recovered. }

\subsection{The  LSZ-factor}
\label{sec:LSZ}

The LSZ-factor \eqref{eq:JBp} can be extracted from the following diagonal correlation function 
 \begin{eqnarray}
 \label{eq:C}
C(p_B^2, \momB^2) &=& i  \int_x e^{i x \rr}  \matel{ \phiB( \momB) }{ T \Hpz(x)  \Hpzda(0)  }{ \phiB( \momB)}   \nonumber \\
&=&  \int \frac{ds}{2\pi i} \frac{  \text{disc}_s C(s, m_B^2   ) }{s-p_B^2-i0} =   \frac{ |\ZBp|^2    }{ m_B^2- 
p_B^2} + \dots \;,
 \end{eqnarray}
where \eqref{eq:onshell} has been assumed and  as before  the dots stand for higher states.  The  LSZ-factor is formally given by
$|\ZBp|^2  =   \lim_{p_B^2 \to m_B^2} (  m_B^2- p_B^2) C(p_B^2, m_B^2) $ {and translates to} 
\begin{equation}
\label{eq:cZB}
|\ZBp|^2 =  \int_{m_+^2}^{s_0} ds  e^{\frac{(m_B^2-s)}{M^2}} \rho_C(s) \;,
\end{equation}
for the sum rule procedure.
As previously  $2 \pi  i \rho_C(s) = 2i \Im C(s,m_B^2) =  \text{disc}_s C(s, m_B^2   )$ and  $M^2$ and $s_0$ are the Borel parameter and continuum threshold. The analogous breakdown of the spectral density  reads
\begin{equation}
\rho_C(s) = \rho^{(0)}_{C,\mathbb{1}}(s) + 
 \VEV{\bar qq} \rho^{(0)}_{C,\bar qq}(s)  + \frac{\al}{\pi}  ( \rho^{(2)}_{C,\mathbb{1}}(s)+ 
 \VEV{\bar qq} \rho^{(2)}_{C,\bar qq}(s)) \;,
\end{equation}
and  implicitly depends on the further variables $\{ m_B^2, m_b, m_q \}$ 
and the usual renormalisation scale(s).  In the QCD limit, when $\phiB$ is removed, one recovers 
 the standard sum rule for $f_B$ with $|\gphi|^2 \rightarrow m_B^4 f_B^2$. The corresponding Feynman diagrams are given in \FIG\ref{fig:dia-denom}.

\section{The computation}
\label{sec:Comp}

The computation consists of the leading order (LO), real emission and the virtual contribution. 
The most labour intensive part is computing the two loop virtual diagrams, applying the 
Cutkosky rules \cite{Cutkosky:1960sp} in (dimensional regularisation  (DR) $d = 4-2\epsilon$) to obtain the spectral densities. 
The scales present are $m_b, m_q, m_{\ell}$ and $m_B$  and their substitution  allows for different flavour transitions. 
 Below we will present the computation for the main process $\Pi$ \eqref{eq:mainSR} and the 
 denominator $C$ \eqref{eq:C} in parallel. The denominator is simpler to calculate as it is more symmetric and free of leptons.  The absence of leptons mean that we do not need to track collinear logs, $\ln m_\ell$. For $\Pi$  there are soft divergences and collinear logs for which a two cut-off phase space slicing technique is employed.

\subsection{Leading order}

The tree level result for the  \VA  correlator is
\begin{equation}
\label{eq:Pi0}
    \Im \Pi^{(0)} (p_B^2) = \frac{m_+^2  N_c}{8\pi} \, \geff m_{\ell} \bar u \Gamma v \, \frac{\KallenpB^{\frac{1}{2}}}{p_B^2} \bigg( 1- \frac{m_-^2}{p_B^2} \bigg) \mathcal{S}(\epsilon) \; ,
\end{equation}
where the zero superscripts denotes  $\mathcal{O}(e^0)$ and $\mathcal{S}(\epsilon) =  1 - \epsilon \ln ( \frac{\KallenpB}{4\mu^2 p_B^2}) + \epsilon (2-\gamma_E + \ln \pi)$ gives the $\mathcal{O}(\epsilon)$ correction 
($\gamma_E$ is the Euler-Mascheroni constant). Both quark propagators are cut generating a K\"all\'en function of the 
type  
$\KallenpB = \lambda(p_B^2,m_b^2,m_q^2)$.  Similarly the LO denominator correlator result is
\begin{equation}
    \Im C^{(0)} (p_B^2) = \frac{m_+^2 N_c}{8\pi} \KallenpB^{\frac{1}{2}} \bigg( 1- \frac{m_-^2}{p_B^2} \bigg)\mathcal{S}(\epsilon) \, .
\end{equation}
The two results above are similar, not by chance, since some powers of the decay constant will have to cancel.

\subsection{Virtual corrections}
\label{sec:virComp}

\begin{centering}
\begin{figure}
\begin{overpic}[width=1.0\linewidth]{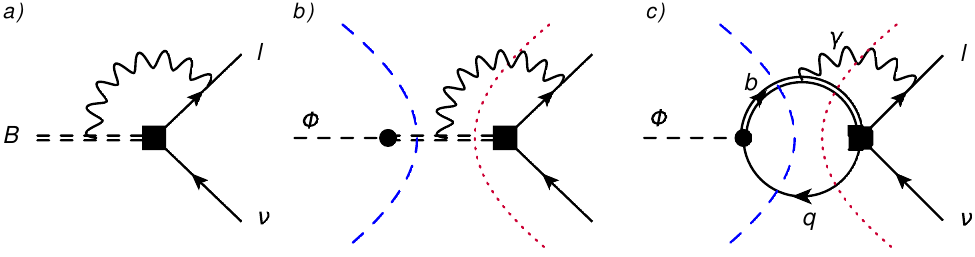}
\put(5,30){Scalar QED}
\put(38,30){Toy Model}
\put(72,30){Structure dependent}
\end{overpic}
	\caption{\small Virtual diagrams for the decay of a $ B^-$-meson in increasing levels of complexity. Left (a): the decay in scalar QED where the meson is point-like. In this case there are also contact diagrams (not drawn) where the photon emerges from the weak vertex (black square). Middle (b): a toy model (cf. \APP\ref{app:toy}) where we make use of the $\Hpz$ operator (black circle) but we keep the meson point-like. Right (c): the full calculation, i.e. the $b \, \ell_1$ contribution to $\Pi(p_B^2,p_\Phi^2)$. For (b) and (c) where there is a dispersion, the two classes of cuts in $p_B^2$ are shown. Those where the photon is not cut are what we refer to as blue cuts (dashed), while those where it is are the red cuts (dotted).}
	\label{fig:redbluecut}
\end{figure}
\end{centering}

When using Cutkosky rules to calculate the imaginary parts of the virtual diagrams the cuts fall into 
two classes cf. \FIG \ref{fig:redbluecut}. 
Ones where the photon is not cut but the quarks are, which we refer to as the ``blue cuts", and ones where the photon 
is cut in addition, which we refer to as the ``red cuts".\footnote{Now, in the real emission case \cite{Janowski:2021yvz} only the blue cuts need to be taken into account when the photon is energetic. 
In this work we consider soft photons as well which is in any case forced upon us by the virtual diagrams 
which do not discriminate between hard and soft momenta.}  In the red cuts, as the photon is put on shell, the calculation resembles a real decay. While for the blue cuts the calculation is simplified as the photon loop factorises.

Below we will go through the procedure taking the $\bl$-diagram as an example. The 
Feynman integral representation of the main process assumes the form 
\begin{multline}
        \Pi^{(2)}_{\bl} = \frac{- m_+ \geff e^2 Q_b Q_{l_1} N_c \mu^{4\epsilon}}{(2\pi)^{2d}} \int \dd[d]{k} \dd[d]{l} \\ \frac{ \Tr \small[ \gamma_5 (-\slashed{l} +m_q)\hat{\Gamma}_\mu (\slashed{q} - \slashed{l} + m_b) \gamma_\rho(\slashed{p_B} - \slashed{l} + m_b) \small] }{(l^2 - m_q^2) ( (p_B-l)^2-m_b^2 ) ( (q-l)^2-m_b^2 ) k^2}
     \frac{\bar u \gamma_\rho (\slashed{l_1}-\slashed{k}+m_{\ell}) \Gamma_\mu v}{(l_1-k)^2-m_{\ell}^2} \; ,
\end{multline}
where we have chosen the Feynman gauge for brevity in the presentation (gauge invariance of the procedure 
is discussed in \SEC\ref{sec:gauge}). 
Taking a blue cut then involves cutting the $l$ and $p_B-l$ momentum propagators 
and this leads to the form 
\begin{multline}
       \Im \Pi^{(2)}_{\bl}|_{\text{blue}} = \frac{-2i \pi^2 m_+ \geff e^2 Q_b Q_{l_1} N_c \mu^{4\epsilon}}{(2\pi)^{2d}} \int \dd[d]{l} \delta^+(l^2 - m_q^2) \delta^+((p_B-l)^2-m_b^2) \\ \times \int\dd[d]{k} \frac{\Tr[ \cdots ] \bar u \cdots v}{( (q-l)^2-m_b^2 ) k^2 (l_1-k)^2-m_{\ell}^2} \;,
\end{multline}
where the trace and spinor structures are dot-suppressed for brevity. 
The photon loop  (in $d^d k$) factorises and can be calculated straightforwardly.  
The integration over $d^d l$ is less standard since one has to take into account 
the delta functions $\delta^+(p^2 - m^2) \equiv \theta(p^0) \delta(p^2-m^2)$ which force the quarks on-shell. 
In the case of the red cuts  the $l$,  $q-l$ and $k$ propagators are cut which leads to the following form
\begin{multline}
\label{eq:bl1Red}
       \Im \Pi^{(2)}_{\bl}|_{\text{red}}  = \frac{-4\pi^3 m_+ \geff e^2 Q_b Q_{l_1} N_c \mu^{4\epsilon}}{(2\pi)^{2d}} \int \dd[d]{k} \frac{\delta^+(k^2)}{(l_1-k)^2-m_{\ell}^2} \\ \times \int \dd[d]{l} \delta^+(l^2 - m_q^2) \delta^+((q-l)^2-m_b^2) \frac{\Tr[ \cdots ] \bar u \cdots v}{(p_B-l)^2-m_b^2} \, .
\end{multline}
As noted earlier, the $\delta^+(k^2)$-function  enforces the photon to be on-shell and makes this resemble a real decay. 
The integrals have at most two open Lorentz indices due to the contraction with 
the spinor structures. The angular integration is generally quite complicated due to the many mass scales involved and the need to retain all $\mathcal{O}(\epsilon)$-terms because the energy integrals contain soft divergences in $1/\eps$.
Fortunately, all integrals required  are known in the literature \cite{Harris:1995tu,Beenakker:1988bq,Somogyi:2011ir}.
The  calculation is simplified by choosing the  $p_B$ rest frame  and exploiting 
systematically the symmetry  between the $b$ and $q$ quark where possible. 
In the case of the $bb$- and $qq$-diagrams (in the main process and for the LSZ-factor) there is an additional 
technical difficulty in terms  double propagators: $1/(l^2-m_q^2)^2$. When cut this generates derivatives of delta functions, $2\pi i \delta'(l^2-m_q^2)$, which can be   dealt with by 
\begin{equation}
\label{eq:Ntrick}
    \Im \int \frac{1}{(l^2-m_q^2)^2} = \lim_{N^2 \rightarrow m_q^2} \dv{N^2} \Im \int \frac{1}{l^2-N^2} \; ,
\end{equation}
commuting the $\Im$ and derivative operations. The variable $N$ is kept distinct from $m_q$ in order to avoid 
differentiation of other $m_q$-dependent parts. 
The three diagrams involving leptons are complicated by $\epsilon_{\mu \nu \rho \sigma}$ structures contracted with spinors. These can be removed via the Chisholm identity in $d=4$ only after we have removed all poles. Using the delta functions, the diagrams are calculated analytically up until a final  {single} numerical integral. All the diagrams have been checked numerically to per mille precision using \textsc{pySecDec}  \cite{Borowka:2017idc} to $\mathcal{O}(\epsilon^0)$.

\subsubsection{{Infrared sensitive (soft) terms}}
\label{sec:IR}

%The heart and life of QED-corrections are the infrared sensitive terms.  
%As discussed in the introduction they consist of the universal soft IR divergences which 
%are well-described by scalar QED and the hard collinear divergences for which are equally 
%described by scalar QED except when there is chirality suppression such as in 
%the leptonic decay with \VA interactions.
Here we aim to discuss infrared divergences which are of two types.  
The universal soft IR divergences due to the massless photon, which are well-described by scalar QED, 
and artificial $\ln m_q$-terms which have to cancel in all observables. 
The physical collinear logs $\ln m_\ell$ originating from  lepton diagrams are discussed in \SEC\ref{sec:Coll}.

When taking individual cuts soft and collinear divergences proliferate which would otherwise cancel once cuts are combined.
Let us first consider artificial divergences of the soft and collinear type. The former are regulated by $1/\epsIR$ and the latter, due to the presence of an internal light quark, take on the form $\ln m_q$. Both of them  
  must and do cancel (between  red and blue cuts) in each individual diagram. 
 This shows the importance of taking into account all possible cuts, discussed further in \APP\ref{sec:allCuts}.  
 
 Let us turn to the physical IR sensitive terms. 
As is well-known, soft divergences are generated by at least two on-shell particles and 
thus in the main process the $\Phi\,l_1$-diagram is the relevant diagram. 
 It contains a soft divergence which only cancels against the 
 real radiation and must be isolated as a $1/\epsIR$ pole. 
 The other main process diagrams only have artificial soft divergences that cancel between  red and blue cuts.

The $\ln m_q$, $\ln m_\ell$ and $1/\epsIR$ poles arise in a different manner depending on the type of cut. 
For the blue cuts they are generated straightforwardly from the factorised photon one-loop integration.  The red cuts are technically  more demanding as 
they cut through both loops. We adapt the phase space slicing method for the treatment of real radiation 
(e.g. \cite{Harris:2001sx}) to the virtual diagrams. 
Taking again the $\bl$-example in \eqref{eq:bl1Red} and letting the result of the $l$-integral be $I_{\bl}$, we obtain 
\begin{align}
    \Im \Pi^{(2)}_{\bl} |_{\text{red}}    = \, &\frac{\pi^3 m_+ \geff e^2 Q_b Q_{\ell_1} N_c \mu^{4\epsilon}}{(2\pi)^{2d}} \int \frac{\dd[d-1]{\bm{k}}}{|\bm{k}|} \frac{I_{\bl}}{k\cdot l_1} \theta(q^2-m_+^2) \nonumber \\ 
    = \, &\frac{\pi^3 m_+ \geff e^2 Q_b Q_{\ell_1} N_c \mu^{4\epsilon}}{(2\pi)^{2d}} \int_0^\Lambda \dd{|\bm{k}|} \int \dd{\Omega_{d-1}} \frac{|\bm{k}|^{d-3}}{k \cdot l_1} I_{\bl} \, .
\end{align}
The $\theta$ function generated by the $l$-integral acts to cut-off the photon energy at
\begin{equation}
\label{eq:La}
\Lambda \equiv \frac{p_B^2 - m_+^2}{ 2\,  (p_B^2)^{1/2}} \;.
\end{equation}
{The $k \cdot l_1$ denominator gives the $\ln m_\ell$-terms when $\bm{k}$ and $\bm{l}_1$ are parallel}, while there is a soft divergence since $I_{\bl} \sim 1/|\bm{k}|$ for $|\bm{k}|\rightarrow 0$. In the slicing procedure we decompose the photon energy integral into a soft part, treated in $d = 4-2\epsilon$, and a hard part in $d=4$
\begin{equation}
    \int_0^\Lambda \dd{|\bm{k}|} \int \dd{\Omega_{d-1}} \to  \int_0^{\Delta E_s} \dd{|\bm{k}|} \int \dd{\Omega_{d-1}} + \int_{\Delta E_s}^\Lambda \dd{|\bm{k}|}  \int \dd{\Omega_{3}} \;.
\end{equation}
The boundary between the two regions is $\Delta E_s$ , and   chosen to be sufficiently small to  drop subleading powers of $|\bm{k}|$ in the soft integrand. The hard part is  to be further decomposed into collinear and non-collinear sections (cf.  \SEC \ref{sec:virColl}). 
%The soft divergence is regulated by $\epsilon$ and care must be taken to track finite terms generated from multiplication of the soft pole with $\mathcal{O}(\epsilon)$ contributions from angular integrations.
  
In order to illustrate the soft pole we have to turn to the  $\Phi\,\ell_1$-diagram 
\begin{equation}
    \Im \Pi_{\Phi\, \ell_1}^{(2)} = \frac{\alpha}{4\pi} Q_\Phi Q_{\ell_1} \frac{m_B^2+m_\ell^2}{m_B^2-m_\ell^2}\ln ( \frac{m_B^2}{m_\ell^2} ) \frac{\Im \Pi^{(0)}(p_B^2) }{\epsIR} + \mathcal{O}(\epsilon^0) \;,
\end{equation}
which matches the universal soft results \cite{Yennie:1961ad,Weinberg:1965nx} 
and cancels against  the real radiation after  Borel transformation.  Note that the expression above contains both a soft pole and a collinear log. There are additional soft-collinear terms $\ln^2 m_\ell$ of the same origin   which 
 cancel similarly. Note that the separation of the soft region is frame dependent due to the $\Delta E_s$ (defined in the $p_B$ rest frame) but since 
the final results are finite in the  $\Delta E_s \rightarrow 0$ limit this is of no concern.

In the denominator the only soft divergent one-particle irreducible (1PI) diagram is the $\Phi\,\Phi$-diagrams as it  is the only  diagram 
without a blue cut
\begin{equation}
    \Im C^{(2)}_{\Phi\,\Phi} = \frac{\alpha}{2\pi} Q_\Phi^2 \frac{\Im C^{(0)}(p_B^2)}{\epsIR} + \ORD(\epsilon^0) \;.
\end{equation}
IR-finiteness is achieved by taking into account  the  $\Phi$ self energy in the on-shell scheme (cf. \SEC\ref{sec:ren}).

\subsubsection{Gauge invariance}
\label{sec:gauge}

 In the explicit computation the Feynman gauge, $\xi = 1$, 
was assumed. Here we turn to the question of gauge invariance by tracking  the 
$(\xi - 1) k_\mu k_\nu / (k^2)^2$-terms.  Whereas these have $1/k^4$-parts, they are 
 simplified as Ward identities contract  propagators to a point. The blue cuts vanish separately for each diagram in DR  while the red cuts cancel by charge conservation $Q_\Phi = Q_{\ell_1} = Q_b - Q_q$. Explicitly, gauge invariance  for the (1PI) diagrams reads
\begin{align}
\label{eq:NumGI}
        \Im \Pi^{(2)}|_{1-\xi} = & \, \frac{4 \pi^4 m_+^2 \geff e^2 N_c \mu^{4\epsilon}}{(2\pi)^{2d}} \bigg( Q_{\ell_1} ( Q_\Phi - Q_b + Q_q ) - Q_\Phi ( Q_b - Q_q ) + (Q_b - Q_q)^2 \bigg) \nonumber \\ & \quad \times \lim_{N^2 \rightarrow 0} \dv{N^2} \int \dd[d]{k} \delta^+(k^2-N^2) \theta(q^2 - m_+^2) \, \bar u \slashed{q} \Gamma v \, \frac{\Kallenq^{\frac{1}{2}}}{q^2} \bigg( 1- \frac{m_-^2}{q^2} \bigg) =0  \;,
        \end{align}
{since the bracket of charges vanishes, and equally so for the denominator}
\begin{align}
        \Im C^{(2)}|_{1-\xi} = & \, \frac{2 \pi e^2  \mu^{2\epsilon}}{(2\pi)^{d}} \bigg( Q_\Phi^2 - 2 Q_\Phi ( Q_b - Q_q ) + (Q_b - Q_q)^2 \bigg) \nonumber \\ & \quad \times \lim_{N^2 \rightarrow 0} \dv{N^2} \int \dd[d]{k} \delta^+(k^2-N^2) \theta(q^2 - m_+^2) \Im C^{(0)} (q^2) =   0 \; ,
\end{align}
where $\Kallenq = \lambda(q^2,m_b^2,m_q^2)$. Note that the first  equation has the same structure 
as the LO expression with $p_B^2 \rightarrow q^2$ which is related to the Ward identity statement above.

\subsubsection{Renormalisation}
\label{sec:ren}

{In addition to  IR divergences there are 
 ultraviolet (UV) divergences with some of them being  peculiar to QED.} For example, the $J_B$-operator which is renormalisation group invariant in QCD, 
renormalises in QED due to $Q_b \neq Q_q$.
 In $\overline{\text{MS}}$ {we obtain that}  $\Hpz$ \eqref{eq:JBp}, setting $m_q = 0$,  
  renormalises as\footnote{For convenience, we define $\frac{1}{\hat{\epsilon}} = \frac{1}{\epsilon} -\gamma_E + \ln 4 \pi$.} 
 %\footnote{We credit Saad Nabeebaccus for this calculation. Note multiplicative renormalisation is spoiled when $m_q >0$.}
\begin{align}
    Z_{\Phi_B J_B} = & 1 + \frac{\alpha}{4\pi} \frac{1}{\hat{\epsilon}_{\textrm{UV}}} \bigg( -(3+\xi)Q_b Q_q - \xi Q_\Phi (Q_b - Q_q) + \frac{1}{2} \xi ( Q_b^2 + Q_q^2 ) - \frac{1}{2}(3-\xi) Q_\Phi^2 + 3 Q_b^2 \bigg)  \nonumber \\
    = & 1 + \frac{3\alpha}{8\pi} \frac{1}{\hat{\epsilon}_{\textrm{UV}}} (Q_b^2 - Q_q^2)\;,
\end{align}
where we made use of charge conservation in the second line. 
For $Q_q = Q_b$ one recovers the non-renormalisation of 
$J_B$ in QCD.

Turning to our computation matters are slightly  complicated since we keep 
both  $m_q $ and $m_b$ non-zero and this leads to   non-multiplicative renormalisation.  All 1PI renormalisations are performed in $\overline{\text{MS}}$, except for the $b$-quark mass for which we 
 considered several options but found the kinetic scheme to give the most stable results 
 (cf. \SEC\ref{sec:num}). 
 The self energy (SE) corrections of the external particles are handled, as usual,  in the 
 on-shell scheme  (e.g. \cite{Sterman:1993hfp})
\begin{align}
\label{eq:Zfactors}
    \Im \Pi^{(2)}_{\text{SE}} =& \frac{\alpha}{2\pi} \bigg( Q_\Phi^2 \delta Z_S + Q_{\ell_1}^2 \delta Z_2 \bigg) \Im \Pi^{(0)} (p_B^2)  \;, \nonumber \\
    \Im C^{(2)}_{\text{SE}} = & \frac{\alpha}{\pi} Q_\Phi^2 \delta Z_S \Im C^{(0)}(p_B^2) \; ,
\end{align}
with scalar and fermion $Z$ factors
\begin{align}
    4 \, \delta Z_S =& \, (3-\xi) \bigg( \frac{1}{\hat{\epsilon}_{\textrm{UV}}} - \frac{1}{\hat{\epsilon}_{\textrm{IR}}} \bigg) + 1-\xi  \;, \nonumber \\
    4 \, \delta Z_2 = & \, -\xi \frac{1}{\hat{\epsilon}_{\textrm{UV}}} - (3-\xi) \frac{1}{\hat{\epsilon}_{\textrm{IR}}} + 3 \ln ( \frac{m_\ell^2}{\mu^2} ) - (3+\xi) \;,
\end{align} 
using the expression in \EQ 2.18 in \cite{Isidori:2020acz}.
It is implicitly understood that we adapt the same scheme choice for the main process and the LSZ-factor to enable necessary cancellations to occur.

 Finally there is the issue of the running of the weak 4-Fermi operator in \eqref{eq:weakH}. 
 Some of the QED-corrections  are  absorbed into the definition of $G_F$ (taken from muon decay)  using the
 W-regularisation \cite{Sirlin:1981ie}. As we regulate in DR this procedure must be adapted 
 as done in  \cite{Brod:2008ss,Gorbahn:2022rgl,Bigi:2023cbv}.  It amounts to the replacement $G_F \rightarrow G_F C(\mu)$ with 
\begin{equation}
    C(\mu) = 1 + \frac{\alpha}{2\pi} \bigg( \ln ( \frac{M_Z^2}{\mu^2} ) - \frac{11}{6} \bigg) \, . 
\end{equation}

\subsubsection{Final virtual rate}

After renormalisation, the denominator $C$ is IR and UV finite and $|\gphi|^2$ can be extracted through \eqref{eq:cZB}.  For the main process we must first   square the amplitude, spin sum and integrate over (the trivial) phase space.  Defining $\Im \Pi = \Im \overline{\Pi} \,  \bar u \Gamma v$ and $\mathcal{B}_{M^2} \Pi = \frac{1}{\pi} \int \dd{s} \Im \Pi (s) e^{(m_B^2-s)/M^2}$,  the virtual  rate reads
\begin{multline}
      \Big[ |\gphi|^2 \Gamma_{B \rightarrow \ell \bar \nu} \Big] = \frac{(m_B^2 - m_\ell^2)^2}{4\pi \mu^{2\epsilon}m_B^3} \bigg( \big(\mathcal{B}_{M^2} \overline{\Pi}^{(0)} \big) \big( \mathcal{B}_{M^2} \overline{\Pi}^{(0)} \big)^* \\  + 2 \mathcal{K}(\epsilon) \Re \big(\mathcal{B}_{M^2} \overline{\Pi}^{(2)} \big) \big( \mathcal{B}_{M^2} \overline{\Pi}^{(0)} \big)^*  +  \ORD(\al^2) \bigg) \; ,
\end{multline}
where the first and second  term correspond to the LO- and the NLO-result. The quantity 
$\mathcal{K}(\epsilon) = 1 -2\epsilon \ln \frac{m_B^2 - m_\ell^2}{2\mu m_B} + \epsilon (2-\gamma_E+\ln \pi)$ parametrises  phase space corrections which cancel against the real radiation and so in practise can be ignored.

\subsection{Real radiation}
\label{sec:realComp}

\subsubsection{The amplitude}
\label{sec:realAmp} 

The calculation of the amplitude ${\cal A}( \lepB \ga)$  involves three pieces which 
we may write schematically (cf. \eqref{eq:Areal} for the relation to the amplitude)  as follows
\begin{equation}
\Pi^{\ga}_\rho(p_B^2,q^2) = \Pi^{\ga}_\rho|_{\text{lep}}+ \Pi^{\ga}_\rho|_{\Phi}   + \Pi^{\ga}_\rho|_{B} \;,
\end{equation}
with $\rho$  to be contracted with the photon polarisation tensor, $\epsilon^*_\rho$.  
The first and second term are the leptonic and the $\Phi$ particle contributions. Since both sets of particles are treated as point-like 
they are  straightforward.   The third is non-trivial as it is truly structure-dependent and  for which 
we adapt (cf. below \eqref{eq:below}) the NLO LCSR computation  \cite{Janowski:2021yvz}. In that reference 
 the point-like contributions (contact terms) were separated from the truly structure-dependent part and we may thus write, 
 using the shorthand ($\geff$ defined in \eqref{eq:mom} and $q = p_B-k$)
\begin{equation}
\gtilde \equiv m_+ \geff e s_e \;,
\end{equation}
 in the Lorentz gauge\footnote{The flag $s_e = \pm 1$ corresponds to the convention of 
 the covariant derivative cf. \APP\ref{sec:conventions}. It is important to keep track of it when 
 combining the form factor contribution with the $\Phi$-emission diagram.}$^,$\footnote{The structure-dependent part, is given by $ {\Pi}^{\ga}_\rho|^{\text{struc}}_{B} =   (\gtilde/im_+ m_B) \, \bar u \Gamma^\mu v \big( \Pi^V_\perp  \pperp{\mu\rho}  -  \Pi^V_{\para}  \ppara{\mu\rho} \big) $, 
with  $\Pi^V_{\perp,\para}$  given in  \cite{Janowski:2021yvz}. Note that here (the charged case), the form factors $V_{\perp,\para}$ are negative.}
 \begin{alignat}{2}
 \label{eq:3part}
&  \Pi^{(\ga)}_\rho|_{\text{lep}} &\;=\;&
\gtilde Q_{\ell_1} \Pi_{f_B} (p_B^2) \bigg( \, \bar u \gamma_\rho \Gamma v + \frac{m_\ell}{2l_1 \cdot k} \Big( \, \bar u \gamma_\rho \slashed{k} \Gamma v + 2 (l_1)_\rho \, \bar u \Gamma v \Big)  \bigg) \;, \\[0.1cm]
&  \Pi^{(\ga)}_\rho|_{\Phi} &\;=\;&  - \gtilde Q_\Phi \frac{(p_\Phi)_\rho}{p_\Phi \cdot k} \Pi_{f_B} (q^2) m_\ell \, \bar u \Gamma v \;,   \nonumber \\[0.1cm]
&  \Pi^{(\ga)}_\rho|_{B} &\;=\;& \gtilde Q_\Phi \, \bar u \Gamma^\mu v \bigg( \frac{\Pi_{f_B} (q^2)-\Pi_{f_B} (p_B^2)}{k \cdot p_B} (p_B)_\rho (p_B)_\mu - \Pi_{f_B} (q^2) g_{\rho \mu} \bigg) +   {\Pi}^{(\ga)}_\rho|^{\text{struc}}_{B}  \;,\nonumber
 \end{alignat}
 where we have separated the structure-dependent part which depends on the quark charges in a non-trivial manner.
The correlation function $\Pi_{f_B} (p^2)$ is related to the tree level correlator via $\Pi^{(0)} (p_B^2) = \geff m_+ m_\ell \, \bar u \Gamma v \, \Pi_{f_B} (p_B^2)$ defined in \eqref{eq:Pi0}. 
It is noted that the function arguments differ in the $\Phi$- and lepton-case because the photon is radiated before ($q^2$) versus after ($p_B^2$) the $B$-meson is resolved respectively.
Finally, the structure-dependent part is contained in two form factors $V_{\perp,\para}$ defined from 
 \cite{Janowski:2021yvz}
\begin{equation}
\label{eq:matel}
\matel{\gamma}{ \bar q \hat{\Gamma}_\mu b}{B} = - \frac{s_e e}{m_B} \bigg( P_\mu^\perp V_\perp (q^2) - P_\mu^\para ( V_\para (q^2) + Q_{ B^-} \frac{m_B f_B}{k \cdot p_B} ) - P_\mu^{\rm{Low}} Q_{B^-} 
\frac{m_B f_B}{k \cdot p_B} \bigg) \;,
\end{equation}
where $P^\mu \equiv P^{\mu \rho} \, \eps^*_{\rho}(k)$ with 
% $\pperpparalow{\mu}    \equiv \eps^{*\rho} \pperpparalow{\mu \rho }$)
\begin{equation}
\label{eq:P} 
 \pperp{\mu\rho} \equiv   \varepsilon_{\mu \rho \be \ga } (p_B)^\be k^\ga  \;,  \quad    \ppara{\mu \rho }   \equiv  i  \,  (   k \cdot p_B \, g_{\mu \rho} -   k_{\mu} \,\pB{\rho} ) \;, 
 \quad P^{\textrm{Low}}_{\mu\rho} \equiv i  \pB{\mu} \pB{\rho} \;.
\end{equation}
 Assembling all parts we obtain the amplitude (using  \eqref{eq:Areal})
 \begin{alignat}{2}
 \label{eq:3partamp}
&    m_+ (\mathcal{A}^\gamma)^\rho &\;=\;&
{-i \gtilde Q_{\ell_1}}  \FB \bigg( \, \bar u \gamma^\rho \Gamma v + \frac{m_\ell}{2l_1 \cdot k} \Big( \, \bar u \gamma^\rho \slashed{k} \Gamma v + 2 l_1^\rho \, \bar u \Gamma v \Big)  \bigg) +  \frac{\gtilde Q_\Phi  \FB   }{k \cdot p_B}  P_{\textrm{Low}}^{\mu\rho}  \bar u \Gamma_\mu v \, 
 \nonumber \\[0.1cm]
%&   & \;+\;& \gtilde Q_\Phi \, \frac{\bar u \Gamma_\mu v }{\gphi}\bigg( \frac{\mathcal{B}_{M^2}\Pi_{f_B} (q^2)-\mathcal{B}_{M^2}\Pi_{f_B} (p_B^2)}{k \cdot p_B} (p_B)^\rho (p_B)^\mu - \mathcal{B}_{M^2} \Pi_{f_B} (q^2) g^{\rho \mu} \bigg)    \nonumber  \\[0.1cm]
&   &\;-\;&  \frac{\gtilde}{ m_B}  \bigg( V_\perp (q^2) P_\perp^{\mu \rho} - 
\left(V_{\para}(q^2) + \frac{ Q_\Phi m_B \FB}{k \cdot p_B}\right)  P_\para^{\mu \rho} \bigg) \, \bar u \Gamma_\mu v \; , 
  \end{alignat}
where as before non-physical Lorentz structures proportional to $r$ are discarded, and $r$ is 
effectively set to zero everywhere except in $F_B$, the sum rule equivalent of $f_B$ \eqref{eq:naive}, 
\begin{equation}
\label{eq:FB}
\FB  = \FB(M^2) \equiv   \frac{m_+}{\gphi} \mathcal{B}_{M^2} \Pi_{f_B} (p_B^2)  \;,
\end{equation} 
in the sense that $p_B^2 \neq m_B^2$. These terms proportional to $\FB$ must be included in this way in order to guarantee the cancellation of the soft IR divergences. Importantly a term in the $ P_\para^{\mu \rho} $-structure of the type $(\Pi_{f_B} (p_B^2) -\Pi_{f_B} (q^2) )/(p_B^2-q^2)$, required by the equation of motion, has been dropped 
since this term is automatically absorbed into the very definition of $V_\para$ \eqref{eq:matel}, in \cite{Janowski:2021yvz}.
The correspondence between \eqref{eq:matel} and \eqref{eq:3partamp} is established by 
$Q_{\Phi} = Q_{B^-}$  (with the addition of the lepton parts) and for $m_\ell \to 0$ one recovers the well known result that the amplitude is expressed entirely in terms of the form factors. All other terms in  \eqref{eq:3partamp} correspond to the point-like approximation.  
Note that the $V_{\perp} P_{\perp}$ and  $V_{\para} P_{\para}$ -terms are of $\ORD(E_\ga)$ whereas all the remaining terms are of order $\ORD(1/E^\ga)$ or $\ORD((E_\ga)^0)$ which are universal due to Low's theorem. 
Gauge invariance, $k \cdot \mathcal{A}^\gamma=0$, of \eqref{eq:3partamp} follows from applying charge conservation  $Q_\Phi = Q_{\ell_1}$ and $k^\rho P_{\mu \rho}^{\para, \perp} = 0$.

Finally, we consider it worthwhile to comment on the role of what we called blue ($p_B^2$) and red ($q^2$) cuts. 
These cuts start at
\begin{equation}
p_B^2|_{\text{blue}}  > m_+^2  \;, \quad  p_B^2|_{\text{red}} > \big( E_\gamma + \sqrt{E_\gamma^2 + m_+^2} \big)^2  \;,
\end{equation}
respectively. In the LCSR  computation  \cite{Janowski:2021yvz} which is valid at large recoil 
 ($E_\gamma > 1.25 \GeV$ or $q^2 < 14 \GeV^2$), the red cuts never contribute and are simply not necessary \cite{Nabeebaccus:2022jhu}.\footnote{We can make this discussion more concrete by focusing 
 on the Low-type terms (second and third line) in  \eqref{eq:3part}. 
 In the LCSR computation the Low term is  corresponds to 
 $\Pi_{f_B} (p_B^2)(p_B)^\rho (p_B)^\mu$ on the third line. Here however the Low term comes from the second line and the entire third line is subleading in the photon energy (since $p_B^2 \approx q^2$).}

\subsubsection{Final radiative rate}

To infer the rate $\Gamma^{B \rightarrow \ell \bar \nu \gamma}$, the correlator must be Borel transformed,
leading to $i \mathcal{A}^\gamma \gphi$, which is then squared and integrated over the three body phase space. 
Note that only $\Pi_{f_B}$ and not the Lorentz structures are Borel transformed in line with the dispersion relations. In the Lorentz structures $p_B^2 \to m_B^2$ is assumed. The phase space integration for the real rate reads
\begin{alignat}{2}
 &     \Big[ |\gphi|^2 \Gamma_{B \rightarrow \ell \bar \nu \gamma} \Big] &\;=\;& \frac{(2\pi)^d}{2m_B} \int \Big[ \langle |\mathcal{A}^\gamma |^2 \rangle | \gphi |^2 \Big] \delta^{(d)}(p_B-l_1-l_2-k) \\[0.1cm] 
 &   & & \quad \times  \frac{\dd[d-1]{\bm{l_1}}}{(2\pi)^{d-1} 2E_{\ell_1}} \frac{\dd[d-1]{\bm{l_2}}}{(2\pi)^{d-1} 2E_{l_2}} \frac{\dd[d-1]{\bm{k}}}{(2\pi)^{d-1} 2E_{\gamma}} \nonumber  \;,
\end{alignat}
where $\langle \cdots \rangle$ indicates that spin and polarisation sums have been performed. Again, it is convenient to work in the $p_B$ 
rest frame. The soft part of the integral is treated with a conventional real slicing technique to obtain
\begin{alignat}{2}
\label{eq:realSoft}
 &   \Big[  \Gamma_{B \rightarrow \ell \bar \nu \gamma} \Big]_{\text{soft}} &\;=\;& - \frac{\gtilde^2 m_\ell^2 (m_B^2 - m_\ell^2)^2}{16\pi^3 m_B^3 m_+^2 \mu^{2\epsilon}}   \FB^2 \,
     \mathcal{K}(\epsilon) \bigg[ \frac{1}{2} Q_{\ell_1}^2 \frac{m_B^2+m_\ell^2}{m_B^2 - m_\ell^2} \ln \frac{m_\ell^2}{m_B^2}      \\[0.1cm] 
 &  &\;+\;&   \bigg( \frac{-1}{2 \hat{\epsilon}_{\textrm{IR}}} + \ln \frac{m_B \omega_s^\gamma}{\mu} \bigg) \bigg( Q_\Phi^2 + Q_{\ell_1}^2 + Q_\Phi Q_{\ell_1} \frac{m_B^2+m_\ell^2}{m_B^2 - m_\ell^2} \ln  \frac{m_\ell^2}{m_B^2}  \bigg)  + \ldots \bigg] \nonumber  \; ,
\end{alignat}
with $F_B$ as in \eqref{eq:FB}.
The dots stand for finite contributions or soft collinear terms which cancel with the virtual. Notice that there is a finite {$Q_{\ell}^2 \ln m_\ell$-term} generated from this soft region, shown explicitly in \eqref{eq:realSoft}. 
  Again the  $1/\epsIR$ pole is universal and cancels with the virtual provided that the same sum rule specific parameters $s_0, M^2$ are chosen. The soft cut-off $\omega_s^\gamma \equiv 2 ( \Delta E^\gamma_{s} ) / m_B$ dependence cancels against 
the hard region. The hard integration (for which one can assume $d=4$) is easily performed using the Dalitz variables
\begin{equation}
\label{eq:xy}
    x = 1- \frac{(p_B-k)^2}{m_B^2} \;, \qquad y = 1 - \frac{(p_B-l_1)^2}{m_B^2} \; ,
\end{equation}
where $x$ and $y$  serve as dimensionless photon energy and  angular integration variables.
Explicitly the hard integration assumes the form 
\begin{equation}
\label{eq:realHardInt}
    \Big[ |\gphi|^2 \Gamma_{B \rightarrow \ell \bar \nu \gamma} \Big]_{\text{hard}} = \frac{m_B}{256\pi^3} \int_{\omega_s^\gamma}^{r_E} \dd{x} \int_{1-x+\frac{x m_\ell^2}{(1-x)m_B^2}}^1 \dd{y} \Big[ \langle |\mathcal{A}^\gamma |^2 \rangle | \gphi |^2 \Big] \;.
\end{equation}
 The photon energy cut-off is parameterised by  $r_E \equiv 2 \decut / m_B  \in [0, 1-m_\ell^2/m_B^2]$. (Hard) collinear divergences arising from the angular integration over the photon-lepton angle $c_\gamma = \cos \theta$ (defined in the $B$ rest frame) are discussed in the next section.

\section{Collinear logs}
\label{sec:Coll}

We now turn to one of the central conceptual  points of the paper and that is the structure-dependent 
collinear logs $\ln m_\ell$. As discussed in the introduction, their appearance for \VA interactions 
in leptonic decays is rather the exception than the rule as it is owed to the chiral suppression of the 
tree-level amplitude.  For example, the computation of the $B_s \to \mu^+\mu^-$ virtual structure-dependent
corrections in SCET  show such logs  (cf. \EQ 5 in \cite{BBS17}) but what seems not to have been appreciated is that they are specific to that decay with \VA interactions.  These logs must be absent in Yukawa-type \SP interactions as shown by our explicit computation below. 

\subsection{Universal IR logs from the splitting function}
 \label{sec:SP}
 
We consider it worthwhile to illustrate the universality of the collinear logs in the case where the KLN-theorem applies. The point is that the KLN-theorem is based on unitarity which in 
turn implies  that collinear logs $\ln m_\ell$ must cancel between 
real and  virtual parts.  In that case they are reproducible from the splitting function and the LO decay rate e.g.   
\cite{Zwicky:2021olr}. 
This has been exploited  to infer the QED-effects on the charmonium resonances in the low 
$q^2 = (l_+ + l_-)^2$ region in $B \to K \ell^+ \ell^-$  \cite{Isidori:2022bzw}.  The point is that in the SM $B \to K \ell^+ \ell^-$,  unlike the leptonic decay, is not chirally suppressed.  Hence, the non-chirally suppressed \SP case can be treated in the same way using the splitting function\footnote{The $1/(1-z)_+$ is the plus distribution, defined under the integral as         $\int_0^1 \dd{z} f(z)/(1-z)_+ \equiv \int_0^1 \dd{z} (f(z)-f(1))/(1-z)$, which regulates the soft region $z\rightarrow 1$.}
 \begin{equation}
 \label{eq:P}
 P_{f \to f \ga}(z)  =   \frac{1 + z^2}{(1- z)_+}  + \frac{3}{2} \de(1-z) \;,
 \end{equation}
where  ({cf. \eqref{eq:xy}}) 
\begin{equation} 
\label{eq:zx}
z = 1-x  = \frac{(p_B-k)^2}{m_B^2} \;,
\end{equation}
takes the value $z=1$ when the photon is soft. 
The result yields \cite{Zwicky:2021olr}
 \begin{alignat}{2}
 \label{eq:collin}
& \Gamma^{\textrm{NLO}}_{\text{\SP}}  |_{\ln m_\ell} &\;=\;&  -  \frac{\al}{\pi} {Q}^2_{\ell^-}    \ln {m}_\ell 
\int^1_{{1-r_E}} d z  P_{f \to f \ga}(z)   \Gamma^{\textrm{LO}}_{\text{\SP}}  
\nonumber \\[0.1cm]
  &  &\;=\;&   - \frac{\al}{\pi} Q^2_{\ell^-}    \ln{m}_\ell 
 \left( \frac{3}{2} - r_E (2 - \frac{1}{2} r_E ) \right)  \Gamma^{\textrm{LO}}_{\text{\SP}} \;,
 \end{alignat}
 for the collinear logs with photon energies up to the energy cut-off $r_E$. When  fully inclusive in the photon energy, $r_E \rightarrow 1$, the logs cancel completely. This is a rather simple example as for more involved kinematics 
 such as $B \to K \ell^+ \ell^-$ the LO rate depends on the collinear momentum fraction $z$ \cite{Isidori:2022bzw}.
Our point is that we expect \eqref{eq:collin} to hold for the \SP case but not the \VA case.  In what follows 
 we will verify this explicitly using the slicing technique.
 % we can extract the logs through collinear slicing of both the virtual and real hard regions as shown in the next two sections. For the $\tau \bar \nu_\tau$ channel this is inappropriate though indeed the notion of a collinear log has gone.

\subsection{The virtual contribution}
\label{sec:virColl}

In general, hard-collinear (hc) logs can be generated from both the red and blue cuts in the virtual case. 
The blue cuts as emphasised previously are simple since they reduce to one-loop functions. 
 As the red cuts involve an on-shell photon integration, due to the cut photon propagator, the extraction of the collinear part is complicated.
 As a new feature, 
 we adapt the slicing method, from the real radiation, to the virtual collinear case.\footnote{We are not aware as to whether this has been done previously in the literature.  We however have not conducted an extensive search.}
We parameterise the virtual collinear region by $\{z_v, k \cdot l_1 \}$ where
 \begin{equation}
  k = (1-z_v) l_1 \;,
 \end{equation}
 such that $(1-z_v)$ {is proportional to} virtual photon energy. The collinear phase space in the hard regime ($d=4$) becomes
\begin{equation}
    \int_{\text{hc}} \frac{\dd[3]{\bm{k}}}{|\bm{k}|} = 2\pi \int_{\delta_H}^{1-\omega_s} \dd{z_v} \int_{\frac{1}{2}(1-z_v)m_\ell^2}^{ {\omega_c \, p_B^2}} \dd{(l_1 \cdot k)} \; ,
\end{equation}
where 
\begin{equation}
\omega_s \equiv \frac{2 \Delta E_s }{ (p_B^2)^{1/2}} \;, \quad  \delta_H \equiv \frac{m_+^2}{p_B^2} \;,
\end{equation}
are the dimensionless soft  (removing the universal soft-region) and 
   hard  (the dimensionless equivalent of $\Lambda$ \eqref{eq:La}) cut-offs. 
The collinear logs arise from  $1/ l_1 \cdot k $-terms in the integrand and  
are regulated by $m_\ell^2$ in the lower integration limit.  
% Subleading $\mathcal{O}(m_\ell^2)$-terms in the numerator are neglected.   
The upper limit $\omega_c \ll \omega_s$ is a dimensionless collinear cut-off which will cancel against the 
hard non-collinear part and is of no special significance. 
We consider first the  \SP and then the \VA case. 
Comparison to \eqref{eq:collin} requires the inclusion of the real part and  postponed to the next section.

\paragraph{The \SP case:}   The blue cuts are found to match scalar QED,
\begin{align}
    \Im \Pi_{\text{\SP}} (p_B^2)|^{\text{blue}}_{\hc} =
    & - \frac{\alpha}{\pi} Q_{\ell_1} ( Q_b - Q_q ) \Im \Pi^{(0)}_{\text{\SP}} (p_B^2) \ln m_\ell  \nonumber \\ =& - \frac{\alpha}{\pi} \Im \Pi^{(0)}_{\text{\SP}} (p_B^2) \ln m_\ell \; ,
\end{align}
and in fact there cannot be complete cancellation as there are only two diagrams proportional to  $Q_{b}$ and $Q_q$ respectively. This aspect changes for the red cuts which additionally have the  $\Phi\,\ell_1$-diagram (proportional to $Q_\Phi$). Then cancellation can and does occur 
\begin{align}
\label{eq:SPredlogs}
    \Im \Pi_{\text{\SP}} (p_B^2)|^{\text{red}}_{\hc} =& \, \frac{\alpha}{\pi} Q_{\ell_1}Q_{\Hpz} \ln m_\ell  \int_{\delta_H}^{1-\omega_s} \dd{z_v} \frac{z_v}{1-z_v} \Im \Pi^{(0)}_{\text{\SP}} (z_v p_B^2) =  0 \;,
\end{align} 
by  $Q_{\Hpz} =0$ \eqref{eq:Q}.\footnote{Further, since the blue cuts in the hard collinear region 
reproduce all universal $\ln m_\ell$-terms and the red cuts cancel, the same must hold for any hard-collinear logs generated in the soft region. That is, there can be no other hard-collinear logs, and that is what we find explicitly with the same $Q_{\Hpz} =0$ cancellation occurring in the soft region. Since the soft region does not discriminate between \SP and \VA interactions this holds for  \VA as well.} These aspects work very differently in the \VA case.
\paragraph{The \VA case:} 
%We now turn to the \VA case for which there will be structure-dependent logs $\ln m_\ell$. 
The blue cuts read
 \begin{align}
\label{eq:VAbluelogsmq}
    \Im \Pi_{\text{\VA}} (p_B^2)|^{\text{blue}}_{\hc} = & \frac{m_+ \geff \alpha Q_{\ell_1} N_c}{16\pi^2} \frac{m_\ell \bar u \Gamma v}{p_B^4} \bigg(  \nonumber \\ 
    Q_b & \Big[ 2 m_+ m_-^2\KallenpB^{\frac{1}{2}} + (m_+ + m_-)( p_B^2 - m_-(m_+ + m_-) ) p_B^2 L^+ \Big] \nonumber \\ 
    - Q_q & \Big[ 2 m_- (m_+ m_- - p_B^2)\KallenpB^{\frac{1}{2}} + m_- ( m_+ - m_-)^2 p_B^2 L^- \Big] \bigg) \ln m_\ell \; ,
\end{align}
where  
\begin{equation}
    L^{\pm} = \ln \Bigg( \frac{p_B^2 \pm m_+ m_- +\KallenpB^{\frac{1}{2}}}{p_B^2 \pm m_+ m_- -\KallenpB^{\frac{1}{2}}} \Bigg) \; .
\end{equation}
For later convenience, define $L^{\pm}_q$ as $L^{\pm}$ with $p_B^2 \rightarrow q^2$. This time we do not find a result proportional to the LO results.
%since it is not calculable from the splitting function 
%as in the \SP case  \eqref{eq:collin}.
 In the $m_q \rightarrow 0$ limit the result simplifies to
\begin{multline}
    \Im \Pi_{\text{\VA}} (p_B^2)|^{\text{blue}}_{\hc}  = \frac{m_b^2 \geff \alpha Q_{\ell_1} N_c}{8\pi^2} \frac{m_\ell \bar u \Gamma v}{p_B^4} \,  \ln m_\ell  \, \times \\[0.1cm]
     \bigg(  Q_b \Big[ m_b^2 (p_B^2 - m_b^2) + p_B^2 (p_B^2 - 2m_b^2) \ln \frac{p_B^2}{m_b^2} \Big]  + Q_q (p_B^2 - m_b^2)^2  \bigg) \, .
\end{multline}
For the red cuts, the same cancellation by charge occurs for the soft region ($z_v \rightarrow 1$) as in the \SP case, however there \emph{are} extra terms arising from non-soft virtual photons.
\begin{align}
\label{eq:VAredlogs}
      \Im \Pi_{\text{\VA}} (p_B^2)|^{\text{red}}_{\hc}  =& \, \frac{\alpha}{\pi} Q_{\ell_1}  Q_{ \Hpz} \ln m_\ell \int_{\delta_H}^{1-\omega_s} \frac{z_v}{1-z_v} \Im \Pi^{(0)}_{\text{\VA}} (z_v p_B^2)  \dd{z_v} \nonumber \\
     & - \frac{m_+ \geff \alpha Q_{\ell_1} N_c}{8 \pi^2} m_\ell \ln m_\ell \,  \bar u \Gamma v \int_{\delta_H}^{1-\omega_s} f_{\hc}(z_v ,p_B^2)\dd{z_v} \nonumber \\
     = & -\frac{m_+ \geff \alpha Q_{\ell_1} N_c}{8 \pi^2} m_\ell \ln m_\ell  \, \bar u \Gamma v \int_{\delta_H}^{1} f_{\hc}(z_v,p_B^2)\dd{z_v}  \;,
\end{align}
where
\begin{multline}
    f_{\hc}(z_v,p_B^2) = \Kallenq^{\frac{1}{2}} \Big[ (Q_b - Q_q)\frac{m_+ m_-^2 - p_B^2 (m_+ + m_-)}{p_B^4 z_v} + \frac{2Q_b m_- + (Q_b - Q_q)m_+}{p_B^2} \Big] \\
    - \frac{m_-}{2 p_B^2} \Big[ Q_b (m_+ + m_-)^2 L_q^+ + Q_q (m_+ - m_-)^2 L_q^- \Big] + Q_b (m_+ + m_-) (1-z_v)L_q^+ \; ,
\end{multline}
and note that in the collinear region $q^2 = z_v p_B^2$. The function $f_{\hc}$ is not soft divergent and this 
justifies neglecting   $\omega_s$  in \eqref{eq:VAredlogs}. The extra parts beyond the \SP are structure-dependent contributions. The integral may easily be  performed in the $m_q \rightarrow 0$ limit giving
\begin{multline*}
      \Im \Pi_{\text{\VA}} (p_B^2)|^{\text{red}}_{\hc} =  \frac{m_b^2 \geff \alpha Q_{\ell_1} N_c}{16 \pi^2} \frac{m_\ell \bar u \Gamma v}{p_B^4} \,   \ln m_\ell  \times  \nonumber \\[0.1cm]
       \bigg[ (p_B^2 - m_b^2) \Big( 4Q_b(p_B^2 - m_b^2) + Q_q (m_b^2-3p_B^2) \Big) 
     + 2 \Big( Q_b (m_b^4 - p_B^4) + Q_q m_b^2 (2p_B^2 - m_b^2) \Big) \ln \frac{p_B^2}{m_b^2} \bigg] 
   \;.
\end{multline*}
We can then combine this with the result  of the blue cuts, \EQ\eqref{eq:VAbluelogsmq}, which yields the total 1PI hard-collinear logs ($m_q = 0$)
\begin{align}
\label{eq:VAvirtuallogs}
  \Im \Pi_{\text{\VA}} |_{\hc}  =\;& \frac{m_b^2 \geff \alpha Q_{\ell_1} N_c}{16 \pi^2} \frac{m_\ell \bar u \Gamma v}{p_B^4} 
  \, \ln m_\ell  \times \nonumber   \\  & \; \bigg[ (p_B^2 - m_b^2) \Big( p_B^2 (4Q_b - Q_q) - m_b^2 (2Q_b+Q_q) \Big)   + 2m_b^2 (Q_b - Q_q) (m_b^2 - 2 p_B^2 ) \ln \frac{p_B^2}{m_b^2} \bigg] \nonumber \\
    = \;& \, \frac{m_b^2 \geff N_c \alpha}{8 \pi^2} \frac{m_\ell \bar u \Gamma v}{p_B^4} \bigg( p_B^2 (p_B^2 - m_b^2) + m_b^2(m_b^2 -2p_B^2) \ln \frac{p_B^2}{m_b^2} \bigg)\, \ln m_\ell  \; ,
\end{align}
where in the last line we have applied $Q_b = -1/3, Q_q = 2/3$.
The total virtual $\ln m_\ell$ also includes the contribution from the lepton wave function renormalisation ($Z_2$), which from \eqref{eq:Zfactors} is 
$(3\alpha/4\pi) Q_{\ell_1}^2 \Im \Pi^{(0)} \ln m_\ell$. In order to conclude we must analyse the real radiation.

\subsection{The real radiation}
\label{sec:realColl}

In the real case, collinear logs arise from the integration of inverse powers of $k \cdot l_1 = \frac{1}{2}m_B^2 (x+y-1)$ 
over the angular variable $y$ in \eqref{eq:realHardInt}. In the \SP case, where there is no structure-dependent form factors $V_{\para, \perp} (z)$, the angular integration over $y$ can be performed and with  $z$ defined in \eqref{eq:zx} we obtain
\begin{align}
     \Big[ |\gphi|^2 \Gamma_{B \rightarrow \ell \bar \nu \gamma} \Big]^{\text{\SP}}_{\text{hc}} &= - \frac{m_B \gtilde^2 Q_{\ell_1}^2}{16 \pi^3}  \big( \mathcal{B}_{M^2} \Pi_{f_B}^{\text{\SP}} (p_B^2) \big)^2 \int_{1-r_E}^{1-\omega_s^\gamma} \tilde{P}_{f \rightarrow f \gamma} (z) \dd{z} \ln m_\ell \nonumber \\
     &\stackrel{r_E \to 1}{\rightarrow}  \frac{m_B \gtilde^2 Q_{\ell_1}^2}{16 \pi^3}  \big( \mathcal{B}_{M^2} \Pi_{f_B}^{\text{\SP}} (p_B^2) \big)^2 \bigg[ \frac{3}{2} + 2 \ln \omega_s^\gamma \bigg] \ln m_\ell + \mathcal{O}(\omega_s^\gamma) \;,
\end{align}
where  $\tilde{P}_{f \rightarrow f \gamma} (z)$ is the real collinear emission part of the fermion splitting function \eqref{eq:P}  \begin{equation}
    \tilde{P}_{f \rightarrow f \gamma} (z) = \frac{1+z^2}{1-z} \; ,
\end{equation}
with the soft  region ($z \to 1$)  removed (since it is cut off by the upper boundary $1-\omega_s^\gamma$).
%Like in the virtual, the red cuts do not contribute for the \SP hard collinear logs. 
In addition there is  the finite soft log from \eqref{eq:realSoft}. Combining the above hard collinear logs with the soft logs and  virtual corrections, we recover the universal form seen in  \cite{Isidori:2020acz} 
$B \to K \ell \ell$, namely 
\begin{multline}
 \Big[ |\gphi|^2 \Gamma_{B \rightarrow \ell \bar \nu (\gamma)} \Big]^{\text{\SP}}_{\text{hc}} = \frac{\alpha}{\pi} Q_{l_1}^2 \Big[ |\gphi|^2 \Gamma_{B \rightarrow \ell \bar \nu}^{\text{\SP}} \Big]^{(0)}   \,  \ln m_\ell \,  \times  \\[0.1cm]
  \Bigg( \bigg[ \frac{3}{2} + 2 \ln \omega_s^\gamma \bigg]_{\text{real (hard)}} + \bigg[ -1 - 2 \ln \omega_s^\gamma \bigg]_{\text{real (soft)}}  + \bigg[\frac{3}{2} - 2 \bigg]_{\text{virt}} \Bigg) = 0 \;,
\end{multline}
where all logs do and must cancel in the fully-inclusive case.
 Conversely, in the \VA case we get ($\bar z \equiv 1-z)$)
\begin{multline}
        \Big[ |\gphi|^2 \Gamma_{B \rightarrow \ell \bar \nu \gamma} \Big]^{\text{\VA}}_{\hc} = - \frac{m_B \gtilde^2 m_\ell^2 Q_{\ell_1}}{16 \pi^3} \mathcal{B}_{M^2} \Pi_{f_B} (p_B^2) \, \ln m_\ell \, \times  \\[0.1cm]
         \int_{1-r_E}^{1-\omega_s^\gamma} \dd{z} \Bigg[Q_{\ell_1} \tilde{P}_{f \rightarrow f \gamma} (z) \mathcal{B}_{M^2} \Pi_{f_B} (p_B^2) + \frac{m_B}{m_+}  \bar z^2 \, ( {V}_\para ( z \, m_B^2 ) -V_\perp ( z \, m_B^2) ) \gphi  \Bigg]  \;,
\end{multline}
we get besides the splitting function piece (which we could integrate) the form factor parts which truly probes 
the structure of the $B_q$-meson! We note that in the soft limit the \SP and the \VA pieces match since the structure of 
the mesons is not resolved (due to the extra $\bar z^2$-factor). In that case 
it  described by the universal formulae  \cite{Yennie:1961ad,Weinberg:1965nx}.

\section{Numerics and phenomenology}
\label{sec:num}

\subsection{Sum rule numerics}

 The numerical input is given in \TAB\ref{tab:inputParams} in \APP\ref{sec:conventions}. 
 The sum rule procedure requires two  parameters, 
the continuum threshold $s_0$ and the Borel parameter $M^2$, to be fixed.  
By the very assumption of the method, semi-global quark hadron duality (e.g. \cite{Shifman:2000jv}), 
the  threshold $s_0$ ought to be in between the pole 
 $m_B^2$ and the onset of the QCD-continuum around $(m_B + 2 m_\pi)^2$ (or $(m_B+m_\rho)^2$). 
 If the sum rule was exact the value of the Borel parameter is immaterial and thus 
 the  independence thereof is a measure of the quality of the sum rule. 
 In practice the Borel parameter and $s_0$ values  have to be such that      
 the condensate expansion convergences (e.g. NLO terms not exceeding $30\%$) 
 and that the  continuum hadronic contributions  relative to the $B$-pole does not exceed $30\%$ 
 \cite{Shifman:1978bx,Shifman:1978by}. Moreover, 
  we make use of the daughter sum rule (DSR) consistency condition 
\begin{equation}
m_B^2 =   \frac{\int_{m_+^2}^{s_0} s \, \rho_{\Pi^{(\ga)}}(s)  e^{-s/M^2} \dd{s} }{  \int_{m_+^2}^{s_0}  \rho_{\Pi^{(\ga)}}(s) e^{-s/M^2} \dd{s} } \;,
\end{equation}
which must hold in full QCD since $\rho_{\Pi^{(\ga)}}(s) \propto \de(s- m_B^2)$, provided $s_0$ is below 
the QCD-continuum threshold.  Note that to ensure the cancellation of the 
soft IR divergences between real and virtual, $\Pi$ and $\Pi^\gamma$ must 
have the same sum rule parameters. 
On the other hand, the sum rule for the main process $\Pi$ and the  denominator $C$ 
are not required to have the same values. In certain cases they have to be taken rather different 
as otherwise  the above mentioned criteria are not obeyed e.g. \cite{Ball:2004ye}.
 In practice, since QED is rather small the parameters  will be similar to 
the ones of the $f_B$ sum rule.  Following the advertised procedures, we find the following parameters
%For better convergence, we allow the denominator parameters to differ slightly from that of the main process. We find good stability using
\begin{align}
\Pi^{(\gamma)} (p_B^2, p_\Phi^2) :& \qquad s_0 = 37.2(10) \GeV^2 \;,  \qquad M^2 = 7.9(5) \GeV^2 \;, \nonumber \\
C (p_B^2, p_\Phi^2) :& \qquad s_0 = 36.9(10) \GeV^2 \;,  \qquad M^2 = 6.3(5) \GeV^2 \, .
\end{align}
We use the same parameters for all channels $\ell = e , \mu , \tau$ consistent with the 
relative smallness of the QED-correction. 
The most relevant input parameter is  $m_b$ for which the kinetic scheme  \cite{Bigi:1994em}
has proven to 
give good convergence in QED-related sum rule physics (e.g.  $B \to \ga$ form factor \cite{Janowski:2021yvz},
 the $g_{BB^*(B_1)\ga}$ coupling \cite{Pullin:2021ebn} and the neutral-charged mass difference 
 \cite{Rowe:2023jlt}). We have found similar patterns and therefore use the kinetic scheme 
 with ${m_b^{kin}}=4.53(6)\GeV $ at $\mu_{\text{kin}} = 1  \GeV$ (with $\overline{\text{MS}}$-conversion formulae given in \cite{Gambino:2017vkx}).  However, in view of that our computation is 
 LO in $\al_s$ we  vary the $m_b$-uncertainty by $\de m_b = \pm 0.2\GeV$ thereby accommodating 
 the $\overline{\text{MS}}$ or pole scheme. 

%In the ideal case, we would use our real form factors $V_{\perp,\para}$ derived from the cutting rules as described in \SEC\ref{sec:realAmp} for the computation. However, these are LO in QCD and consequently do not adequately resolve the $B^{*,1}$ poles near at the kinematic endpoint $q^2 \rightarrow m_B^2$. To give more accurate results, we instead use the NLO, twist$-1,2$ form factors of \cite{Janowski:2021yvz}. In practise, this amounts to the replacement of the form factors in \eqref{eq:replaceFF}. Finally all other input parameters for our calculation as well as details of the slicing cut-off's are summarised in \APP\ref{sec:conventions}. 

The main result reported is the relative  QED-correction to the LO decay rate, $\Delta \Gamma_{\text{QED}}$,  as  a function of the photon energy cut $\decut$ 
\begin{equation}
\label{eq:deltaQED}
1+ \Delta \Gamma_{\text{QED}} (\decut) \equiv \frac{1}{\Gamma^{(0)}} \Big( \Gamma^{B \rightarrow \ell \bar \nu} + \Gamma^{B \rightarrow \ell \bar \nu \gamma}_{E_\gamma < \decut} \Big)
+ \ORD{(\alpha^2)} \, .
\end{equation}

Uncertainties are computed by varying the input parameters, and adding uncertainties in quadrature.  
It is worthwhile to recall that for structure-dependent corrections, unlike scalar QED, 
the virtual and the real corrections are separately well-defined. Thus, it makes sense to discuss them 
separately. 
The virtual contribution is rather stable showing a $\mathcal{O}(6$-$7\%)$ uncertainty. 
It is mainly due to the $m_b$-variation for which the uncertainty, when integrated over the $s$-integral \eqref{eq:Areal}
from $m_b^2$ to $s_0$, tends to cancel which might be seen as accidental. Instead, taking the absolute value of the uncertainty results in an increase by a factor of two to three.  
In order to remain on the conservative side, we therefore double the original uncertainty.
The virtual structure-dependent corrections are given in \TAB\ref{tab:virt}
where the hierarchy  due to large logs generated by $\ln m_\ell/m_B$ is visible.  
\begin{table}[h]
	\centering
	\begin{tabular}{l |  ccc }
 & $\tau$ & $\mu$ & $e$ \\ \hline  
 $\Delta \Gamma_{\text{QED}}|^{\text{virtual}}_{\text{struc}} $ &  $+2.9(2)\%$  & $+4.6(6)\%$ & $+7.8(10)\%$ 
	\end{tabular}
	\caption{ \small The virtual structure-dependent QED corrections in $ B^- \to \ell^- \bar \nu$ for $\ell = \tau,\mu,e$. The  small value of the electron mass causes  numerical  issues for $m_\ell \lessapprox 50m_e$. The extrapolation from $m_\tau$ down to $50m_e$ is a straight line in $\ln m_\ell$ 
	from which we infer the actual electron case. This results in the relatively large uncertainty. }
	\label{tab:virt}
\end{table}
Numerically the $q\ell_1$-,$b\ell_1$- and $\Phi \ell_1$-diagrams are dominant and the $\ln m_\ell$-terms contribute about $2\%$  and $5\%$ in the muon and electron case 
respectively.  In practice these virtual numbers are extracted from the full real and virtual computation with 
$V_{\perp,\para}^{B \to \ga} \to 0$ (i.e. structure-dependent real radiation) and real and virtual scalar QED subtracted.

Concerning the real radiation, it is relevant to remember that 
in experiment it is the shape of the final state that enters the fit (folded with detector resolution effects). The shape is taken from a Monte-Carlo simulation (e.g. \PHOTOS). 
Since to-date the latter uses scalar QED it is therefore the shape as compared to scalar 
QED which is important.
\begin{centering}
\begin{figure}[t]
\hspace{-1em}
\begin{subfigure}{0.5\linewidth}
\begin{overpic}[width=\linewidth]{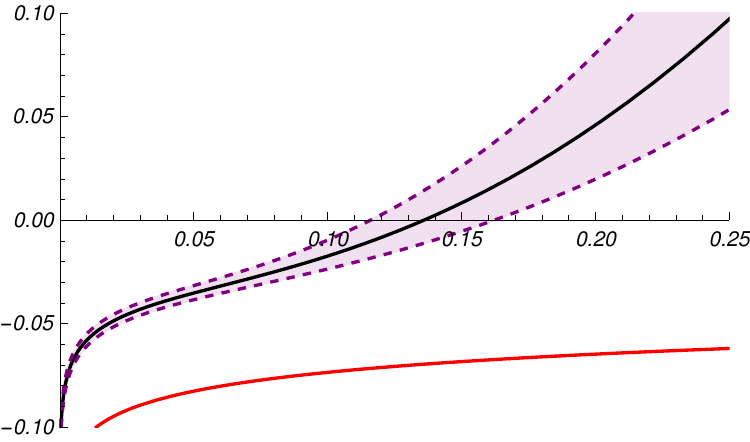}  
\put(33,62){\large $\boxed{B^- \to \mu^- \bar \nu \gamma}$}
%\fboxsep0.4em
%\put(87.75,25.5){\colorbox{white}{\rule[0em]{15pt}{0em} \, }}
%\put(0,53){\colorbox{white}{\rule[0em]{15pt}{0em} \, }}
\put(83,32){\scriptsize $\decut[\text{GeV}$]}
\put(-2,62){${\Delta \Gamma_{\text{QED}}}$}
%\put(60,5){\small {\color{red} scalar QED}}
%\put(34,35){\scriptsize full QED}
\put(43,4){\scriptsize{\color{red} scalar QED}}
\put(43,44){\scriptsize full QED}
\end{overpic}
\end{subfigure}%
\hspace{1em}%
\begin{subfigure}{0.5\linewidth}
\begin{overpic}[width=\linewidth]{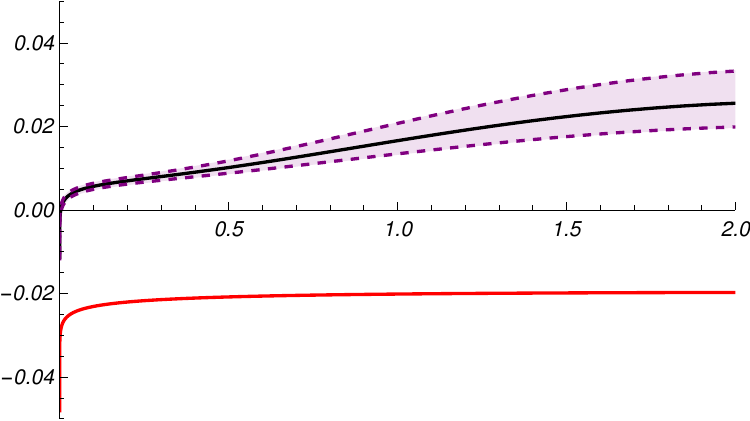}  
\put(33,62){\large $\boxed{B^- \to \tau^- \bar \nu \gamma}$}
%\fboxsep0.4em
%\put(87.5,24){\colorbox{white}{\rule[0em]{20pt}{0em} \, }}
%\put(0,51.5){\colorbox{white}{\rule[0em]{20pt}{0em} \, }}
\put(82,31){\scriptsize $\decut[\text{GeV}]$}
\put(-2,62){  $\Delta \Gamma_{\text{QED}}$}
\put(43,4){\scriptsize{\color{red} scalar QED}}
\put(43,44){\scriptsize full QED}
\end{overpic}
\end{subfigure}
\caption{\small The QED-correction (black line) as a function of the photon energy cut $\decut$ (in $\GeV$), as defined in \eqref{eq:deltaQED}. (Left) The muon channel where there there is a significant rise due to the lifting of helicity-suppression by the structure-dependent terms. (Right) The tau channel where this effect is less significant due to the large $m_\tau$ mass. The error is given by the dashed purple lines and purple bands. For comparison, the red line is the scalar QED result which does not exhibit the same lifting of helicity-suppression. A more detailed version of this plot is given in \FIG\ref{fig:noFF}.}
\label{fig:DeltaPlots}
\end{figure}
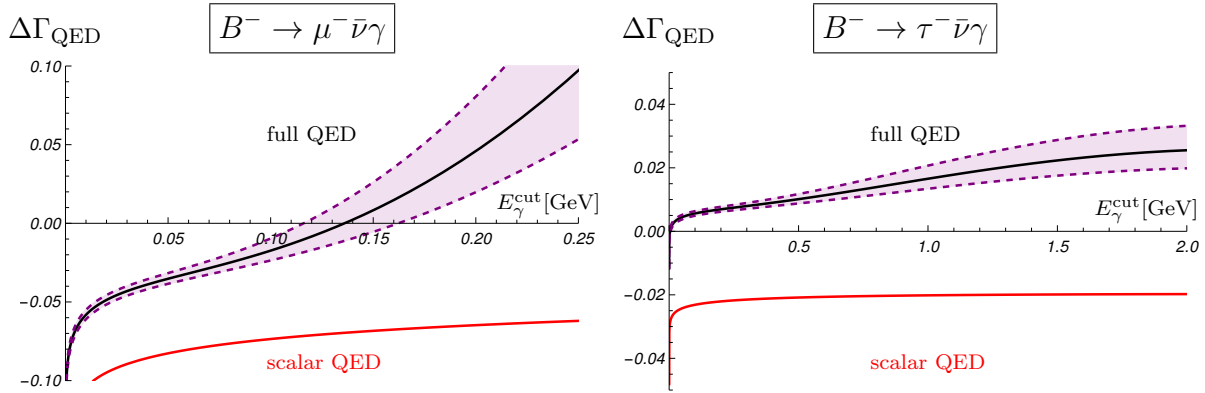
\end{centering} 
In summary not knowing the structure-dependence in effect boils down to an unknown
virtual-correction number (as quoted in \TAB\eqref{tab:virt}) while for the real radiation it is 
the relative shape in $E_\ga$ as compared to the Monte-Carlo which is relevant.
We therefore provide the necessary results in the form of Mathematica notebook as an ancillary file to the
 arXiv version.  
In the remaining paper we follow the tradition and present the correction in terms of a 
photon energy cut-off $\decut$. 
 From a phenomenological viewpoint 
 %For the tau channel, the errors increase from $\sim 11\%$ to $\sim 31\%$ as we get more inclusive in the photon.
the crucial feature about the real structure-dependent QED correction is that the 
helicity-suppression is  relieved (by $V_{\perp,\para}$ form factor contributions) and becomes exceedingly large when $\decut \gg m_\ell$. Since these form factor 
uncertainties are sizeable it makes their  impact even more noticeable for large $\decut$.\footnote{The 
form factor uncertainty in $B \to \ga$ for maximum recoil is $\ORD(10\%)$ but at soft-recoil (soft photon) no direct calculation exists and as previously stated 
the form factor is extrapolated between the pole at $B^*(B_1)$ in the vector and axial channel 
and thus the uncertainty  largely depends   on the input of the pole-reside. The latter is computed from 
a double dispersion sum rule  \cite{Pullin:2021ebn} whose accuracy cannot match the single dispersion sum rule \cite{Janowski:2021yvz} for many reasons.}
We provide plots for the muon and tau case in  \FIG\ref{fig:DeltaPlots}; and  further breakdowns are shown in
\FIG\ref{fig:noFF}.

In the muon case 
the $\decut$-interval is restricted to $ 2 m_\mu$ as beyond this the notion of $\Delta \Gamma_{\text{QED}}$ becomes 
meaningless and one 
 should simply consider  $B  \to  \ell \bar \nu \gamma$ as a decay in its own right.  A general feature is that the scalar QED corrections are negative and that the structure-dependent  corrections are positive making 
 the overall effect smaller.  
  However, we re-emphasise that since the Monte-Carlo programs to date are based on scalar QED 
 in those channels, knowing the corrections thereto is of importance.  We observe:
 \begin{itemize}
 \item [] \emph{Muon case:}  the real structure-dependent corrections exceed the virtual structure-dependent corrections for $\decut  > 0.175(31) \GeV $. 
 It is also observed that the QED-correction changes sign for  $\decut = 0.136(23) \GeV$. Note that $E_\ga \approx 0.2 \GeV$ is a  relevant range in view of a typical $K\mu\mu$-distribution for the LHCb-experiment. 
 \item [] \emph{Tau case:} the real radiation is less important due to the large $\tau$ mass. The overall QED-correction is positive and increases slowly to $+2.60(69)\%$ at the endpoint $\decut = \frac{m_B}{2} (1-(\frac{m_\tau}{m_B})^2)$.
 \end{itemize}

The plots in \FIG\ref{fig:EgDifferentialPlots} give the real differential branching ratio. 
Schematically we divide the rate into three 
  characteristic terms
\begin{equation*}
  \frac{d B( B \rightarrow \ell \bar \nu \gamma)}{dE_\gamma} \propto c_1  f_B^2 \frac{m_\ell^2}{E_\ga} + c_2 f_B \frac{m_\ell^2 E_\gamma^2}{m_B^2}(V_\perp-V_\para)  \ln m_\ell  +
c_3 E_\gamma^3 \big( 1- \frac{2E_\gamma}{m_B} \big)(V_\perp^2+V^2_\para)    + \dots \;,
\end{equation*}
where $c_{1,2,3} = \ORD(G_F^2 m_\ell^0)$ are known coefficients.  These three terms may be associated with
three photon energy regions. The first term dominates for ultrasoft photons and in \FIG\ref{fig:EgDifferentialPlots} the  $\frac{1}{E_\ga}$ behaviour is clearly visible in the $\mu$-case 
but less so for the $\tau$-case since $m_\tau \gg m_\mu$ . For soft photons, the proximity of the $B^{*(1)}$ pole enhances the contribution of the second (structure-dependent) term and other similar terms. Finally for non-soft photons, the rate is completely dominated by structure-dependent effects, given in the third term. This highlights the importance of the form factors in the non-ultrasoft region.
As the  latter are largely determined by the pole residues, as has been emphasised in \cite{Becirevic:2009aq}, 
\begin{equation}
V_{\perp (\para)} = \frac{r^V_{\perp (\para)}}{1-q^2/m_{B^{*(1)}}^2} \; , \qquad r^V_{\perp (\para)} = \frac{m_B}{m_{B^{*(1)}}} f_{B^{*(1)}} g_{B B^{*(1)} \gamma} \;,
\end{equation}
it is worthwhile to briefly discuss them. We note  that the values 
in \cite{Becirevic:2009aq}, resulting from averaging a few models, are qualitatively different 
 \begin{alignat}{5}
 & r^V_{\perp}|_{\mbox{\cite{Pullin:2021ebn}}} &\;\approx\;& -0.30(4)  \;,  \quad & &  r^V_{\para}|_{\mbox{\cite{Pullin:2021ebn}}} &\;\approx\;&-0.16(3)   \;, \quad  & & {r^V_{\perp}}/{r^V_{\para}} \approx 1.9 \;,
 \nonumber \\[0.2cm]
  & r^V_{\perp}|_{\mbox{\cite{Becirevic:2009aq}}}  &\;\approx\;& -0.24(4)  \;,\quad & &  r^V_{\para}|_{\mbox{\cite{Becirevic:2009aq}}} &\;\approx\;&-0.20(5)   \;, \quad & &   {r^V_{\perp}}/{r^V_{\para}} \approx 1.2 \;,
   \end{alignat}
in their ratio and as such will give rise to qualitatively different rates (or Dalitz plots). 

Further plots of the double differential real rate in $E_\gamma$-$\cos(\theta)$ and Dalitz-variables as well as comments on real versus virtual structure-dependence are given in \APP\ref{app:plots}.
\begin{centering}
\begin{figure}
\hspace{-1em}
\begin{subfigure}{0.5\linewidth}
\begin{overpic}[width=\linewidth]{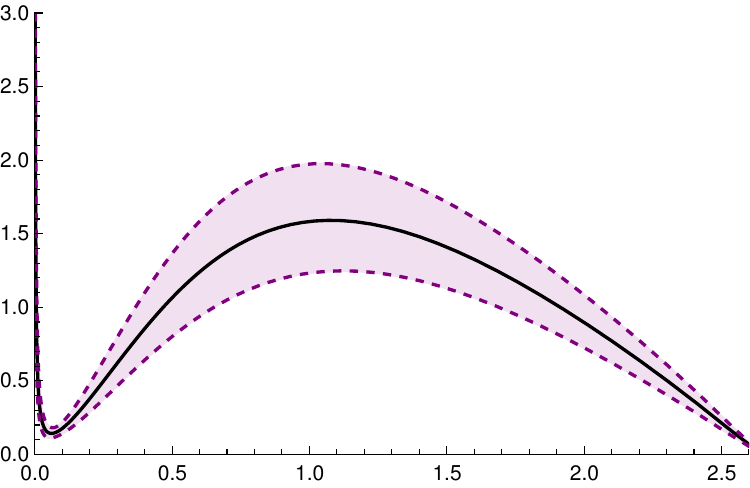}
\put(-10,19){ \small  \rotatebox{90}{$10^6 \dd B/\dd E_\gamma$ $[ \GeV^{-1} ]$ } }
\put(40,-5){ $E_\gamma$ $[\GeV \,]$}
\put(33,65){\large $\boxed{B^- \to \mu^- \bar \nu \gamma}$}
\end{overpic}
\end{subfigure}%
\hspace{2em}%
\begin{subfigure}{0.5\linewidth}
\begin{overpic}[width=\linewidth]{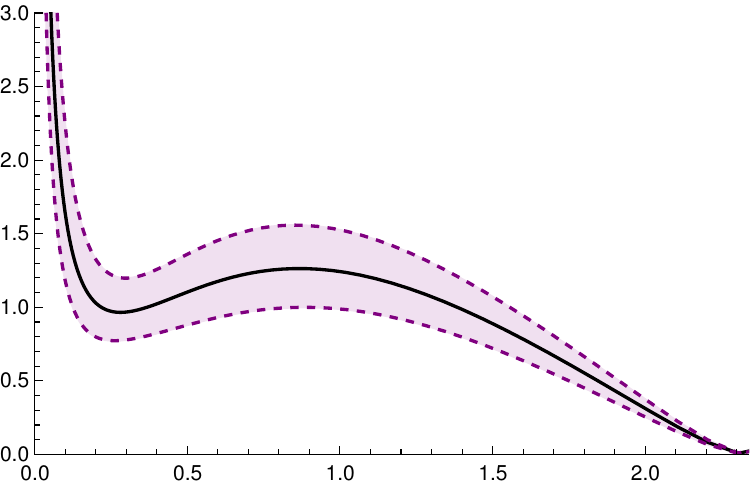}  
\put(-10,19){ \small  \rotatebox{90}{$10^6 \dd B/\dd E_\gamma$ $[ \GeV^{-1} ]$ } }
\put(40,-5){$E_\gamma$ $[\GeV \,]$}
\put(33,65){\large $\boxed{B^- \to \tau^- \bar \nu \gamma}$}
\end{overpic}
\end{subfigure}
\vspace{0.25em}
\caption{\small The differential branching ratio $10^6 \dd{B(B \rightarrow \ell \bar  \nu \gamma)} / \dd{E_\gamma}$ for $\ell = \mu$ (Left) and $\ell = \tau$ (Right) with the uncertainties in purple. Here there is also the uncertainty due to the CKM matrix element $V_{\rm{ub}}$.}
\label{fig:EgDifferentialPlots}
\end{figure}
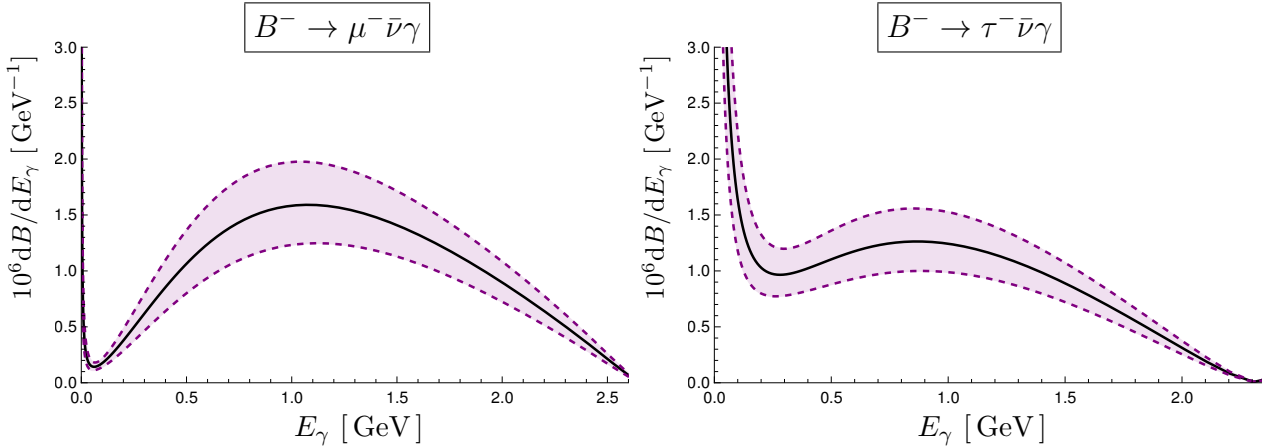
\end{centering}

\subsection{Outlook on phenomenology}
\label{sec:outlook}

Let us turn to the relevance for experimental searches. 
The branching fraction of the leptonic decay without photons and  neglecting the neutrino mass is given by
\begin{equation}
\label{eq:BF}
{\cal B}( B^- \to \ell^- \bar \nu) =  \frac{G_F^2 m_B}{8 \pi} m_\ell^2 \left( 1- \frac{m_\ell^2}{m_B^2} \right)^2 
f_B^2 |V_{\rm ub}|^2 \tau_B \;,
\end{equation}
where $G_F = 1.166 \times 10^{-5} \GeV^{-2}$ is the Fermi constant, $\tau_B = 1.638(4) \times 10^{-12} s$ the $B^+$ lifetime, $|V_{\rm ub}|$ a CKM matrix element 
and $f_B$  the $B$-meson decay constant, $\matel{0}{\bar q \ga_\mu \ga_5 b }{B(p)} = i f_B p_\mu $. 
The SM predictions are  (e.g. \cite{Belle-II:2018jsg})
\begin{equation*}
{\cal B}(B \to e \bar \nu) \approx  8.1(6) \cdot 10^{-12} \;, \quad 
{\cal B}(B \to \mu \bar \nu)  \approx  3.5(3) \cdot 10^{-7} \;, \quad 
{\cal B}(B \to \tau \bar \nu) \approx  7.7(6) \cdot 10^{-5}  \;.
\end{equation*}
The main unknowns in \eqref{eq:BF} are $|V_{\rm ub}|$ and $f_B$ of which the latter is directly calculable by non-perturbative 
methods such as lattice QCD 
and its current uncertainty is  already below a percent \cite{FlavourLatticeAveragingGroupFLAG:2021npn}.  The remaining uncertainty is due to $|V_{\rm ub}|$ 
and is related to the $|V_{\rm ub}|$-$|V_{\rm cb}|$ tension  between inclusive/exclusive-determinations   as 
well as the CKM triangle fit (e.g. \cite{Ricciardi:2021shl}). The $\tau$-rate has been measured at the $B$-factories Belle \cite{Belle:2012egh,Belle:2015odw}
and BaBar \cite{BaBar:2009wmt,BaBar:2013mob}
and its PDG average ${\cal B}(B \to \tau \bar \nu) = 1.09(24) \cdot 10^{-4}$ \cite{PDG} shows some tension with the prediction. However, uncertainties are too large for the QED-correction computed in this paper to be relevant unlike in future measurement (as discussed below).   
For   the $\mu$- and $e$-rate only confidence limits exists. Whereas for  the electron the SM rate is beyond 
the Belle II capabilities the  $\mu$-rate is expected to be measured  with 
$6 \text{ab}^{-1}$ at the $5\sigma$ confidence level   \cite{Belle-II:2018jsg}. 
At $50 \text{ab}^{-1}$  the estimated uncertainties, which are statistically dominated,  
 are \cite{Belle-II:2018jsg}
\begin{equation}
\label{eq:DelBelle}
\Delta {\cal B}(B \to \mu \bar \nu)|_{@50 \text{ab}^{-1}} \approx 7\%  \;,  \quad 
\Delta {\cal B}(B \to \tau \bar \nu)|_{@50 \text{ab}^{-1}} \approx (6,5)\% \;.
\end{equation}
The two numbers in the $\tau$ case are for a hadronic and a semileptonic tag respectively 
and could be averaged as they come from  different data sets.  The fact that the $\mu$-channel is competitive despite lower statistics is owed to it being a clean final state without decaying hadronically. 
We conclude that, in view of the uncertainties in \eqref{eq:DelBelle},
our predictions of $5\%$ and $3\%$ virtual structure-dependent QED-corrections are certainly relevant.  
The relevance of the real radiation depends on two matters. 
Firstly, how inclusive the measurement is, for which the value of the photon energy cut $\decut$ is indicative (cf. \FIG\ref{fig:noFF}). Second and as emphasised previously, how different the shape of the real radiation is 
with respect to scalar QED (and this is used to inform or compare against QED Monte-Carlo programs). 
In view of \FIG\ref{fig:noFF} one has to conclude that it is rather relevant in the $\mu$-channel.

% omit? 
In view of the $|V_{\rm ub}|$-uncertainty it is attractive to consider ratios  \cite{Tanaka:2016ijq,Belle-II:2018jsg}
\begin{alignat}{2}
\label{eq:Rs}
& R_{\rm{pl}} &\;=\;&   \frac{{\cal B}(B^- \to \tau^- \bar \nu)}{{\cal B}(B^- \to \mu^- \bar \nu)}\Big|_{@50 \text{ab}^{-1}}=  222(26)  \;, \nonumber \\[0.1cm] \quad 
& R_{\rm{ps}} &\;=\;&   \frac{\tau_{B^0}}{\tau_{B_-}}   \frac{{\cal B}(B^- \to \tau^- \bar \nu)}{{\cal B}(B^0 \to \pi^+ (e,\mu)^- \nu) }\Big|_{@50 \text{ab}^{-1}}=  0.54(4) \;,
\end{alignat}
where this quantity drops out and the structure-dependent QED-corrections are even more  relevant. 
For $R_{\rm{pl}}$, $f_B$ equally drops and QED probing the hadrons is then the sole hadronic remnant.
The experimental uncertainties, shown in \eqref{eq:Rs} for $50 \text{ab}^{-1}$, 
are however large, since the systematics are of an entirely different origin and thus add.  
{Whereas this method currently seems a viable alternative, it cannot compete once 
the $|V_{\rm ub}|$-uncertainty  drops significantly. However, in that case  it can still serve as a cross-check 
of the actual $|V_{\rm ub}|$-determination itself.}

\section{Conclusions and summary}
\label{sec:con}

In this work we have applied the framework of gauge invariant interpolating operators, proposed in 
\cite{Nabeebaccus:2022jhu}, in the concrete setting of leptonic decays. As all universal IR logs are reproduced 
this constitutes a strong consistency check  of the method. We show that the $B^- \to \ell^- \bar \nu$ 
 mode with \SP (Yukawa-type) interactions exhibit no  structure-dependent hard collinear  logs in accordance 
 with  the  entirely general arguments based on gauge invariance and the KLN-theorem \cite{Isidori:2020acz}. 
However, the Standard Model  leptonic decay is an exception as the chiral suppression circumvents the assumption 
in   \cite{Isidori:2020acz}. 
Hence, these logs can and do arise as confirmed by our explicit results (and reasoned in \cite{Zwicky:2021olr}).

To the best of our knowledge our work provides the first computation of 
structure-dependent QED-corrections in $B^- \to \ell^- \bar \nu$ (the SCET analysis \cite{Cornella:2022ubo} awaits a further study). 
Our main results are summarised in \FIG\ref{fig:DeltaPlots} (and in more detail in 
\FIG\ref{fig:noFF}), showing  virtual and real structure-dependent QED-corrections separately 
on top of the point-like approximation.  The structure-dependent virtual corrections are $+4.6(6)\%$ and $+2.9(2)\%$ 
in the $\mu$- and $\tau$-case.  
Uncertainties due to the real radiation are sizeable. In practice, for the real radiation, what is important is not its size but 
rather its  shape since that is what enters modern experimental analyses.  
As the shape is found to be qualitatively 
different from scalar QED, used in current Monte-Carlo programs, its inclusion is also of relevance. 
(We append a Mathematica notebook to the arXiv version for this purpose.) Partly for  traditional reasons 
we additionally provide plots of the relative QED-corrections as a function of  the photon energy cut-off $\decut$. 
In the $\mu$-channel 
real and virtual structure-dependent corrections are comparable for $\decut|_\mu \approx 0.18(3) \GeV$, whereas in the $\tau$-channel the real radiation is never comparable even when fully inclusive. 
% where $\decut|_{\rm{max}} = \frac{m_B}{2}(1-m_\tau^2/m_B^2) \approx 2.3 \GeV$). 
The origin of this qualitative difference is that the chirality suppression is effectively not there in  
the $\tau$-case. %(in comparison to the $\mu$-case).  

The actual phenomenological importance is discussed in \SEC\ref{sec:outlook}. The size of the corrections are certainly relevant in comparison to the $B^- \to \ell^-\bar \nu$ theory uncertainty which itself is dominated by 
$|V_{\rm ub}|$ since the $B$-meson decay constant $f_B$ is known to  percent-accuracy.  
The $|V_{\rm ub}|$-dependence leads to an interplay with the $|V_{\rm ub}|$-$|V_{\rm cb}|$ puzzle whose resolution is important  in many ways (e.g. in the prediction of  rare $B$ decays). 
% \MR{The manifestly gauge invariant framework applied here provides a reliable approach to probe structure dependent QED-corrections for a wide variety of charged decay channels which in turn may help to bring some clarity to the situation.}
Our framework equally applies to $D$ decays (and possibly $K$ decays) to which we might return in a future work.

\acknowledgments

RZ is supported by an STFC Consolidated Grant, ST/P0000630/1. 
We are grateful to Will Barter and Franz Muheim for discussions.  
In addition we acknowledge Saad Nabeebaccus for useful discussions and comments on the manuscript.   
Many loop computations are performed with the \textsc{FeynCalc} 
package \cite{FeynCalc1,FeynCalc3} and \textsc{Package-X} \cite{Patel:2016fam}.
%thank saad

\appendix

\section{Conventions, inputs and plots}
\label{sec:conventions}

We work with the QED covariant derivative 
\begin{equation}
D_\mu = \partial_\mu + i e s_e Q_f A_\mu \;,
\end{equation}
with $e >0$ (and $s_e = \pm 1$ are frequently chosen conventions), $Q_\Phi = Q_{\ell_1} = -1$, $Q_b = -1/3$, $Q_u = 2/3$ and we take $\epsilon_{0123}=+1$.\footnote{Note that the notation used in  \cite{Isidori:2020acz} with charges s.t. $\sum \hat{Q}_i =0$ is 
not particularly useful anymore since there is now more than one vertex and the quark charges would take 
on different roles in the two vertices (ingoing charges in $\mathcal{L}_w$ versus outgoing in $\Hpz$).}
The particle masses, lifetimes and other input parameters used for our calculation are summarised in \TAB\ref{tab:inputParams}. 

For the slicing procedure, we take $\omega_s = \omega_s^\gamma$ at $s =m_B^2$, and enforce the hierachy $\omega_c / \omega_s \ll 1$ as far as possible. We take $\omega_s^{(e)} = 10^{-3}$, $\omega_s^{(e)} = 100 \cdot \omega_c^{(e)}$, while the heavier muon channel constrains us to take $\omega_s^{(\mu)} = 5 \cdot 10^{-3}$ and $\omega_s^{(\mu)} = 50 \cdot w_c^{(\mu)}$. No slicing is performed for the $\tau$ channel. These choices give good stability. % with the cut-off's.

%Similarly for $D, D_s$ we take $\omega_s^{(e)} = 10^{-3}$, $\omega_s^{(e)} = 100 \cdot \omega_c^{(e)}$ and $\omega_s^{(\mu)} = 3 \cdot 10^{-2}$ and $\omega_s^{(\mu)} = 30 \cdot w_c^{(\mu)}$. These choices in general give good stability, though for $D_{(s)} \rightarrow \mu \bar \nu (\gamma)$ there is an error of a few percent.

\begin{table}[btp]
\centering
\addtolength{\arraycolsep}{3pt}
\renewcommand{\MeV}{\,{\textrm{MeV}}}
\renewcommand{\GeV}{\,{\textrm{GeV}}}
\renewcommand{\arraystretch}{1.3}
\resizebox{0.6\columnwidth}{!}{
\begin{tabular}{c|c|c} 

\multicolumn{3}{c}{$\mbox{ Boson masses~\cite{PDG}}$}\\\hline
$m_{B^-}$ & $M_W$  & $M_Z$  \\\hline
$5.279  \GeV$ &  $80.377 \GeV$ & $91.19 \GeV$    \\\hline
 
 \multicolumn{3}{c}{$\mbox{Lepton masses~\cite{PDG}}$}\\\hline
$m_\tau$ & $m_\mu$  & $ m_e$    \\\hline
$1.777\GeV$ & $105.7 \MeV$ & $0.511 \MeV$  \\\hline

\multicolumn{3}{c}{$\mbox{Quark masses~\cite{PDG}}$}\\\hline
${m_b^{kin}}_{\lscale{1}}$ & $\bar m_{u\lscale{2}} $ & \\\hline
 $4.53(6)\GeV $ & $2.16^{+0.49}_{-0.26} \MeV $ & \\\hline

\multicolumn{3}{c}{$\mbox{Miscellaneous~\cite{PDG}}$}\\\hline

$\tau_{B^-}$ & $G_F$ & $\alpha^{-1} $  \\\hline
$1.638(4) \times 10^{-12} \textrm{ s}$ & $1.166 \times 10^{-5} \GeV^{-2}$ & $137.036$ \\ \hline
$\qbarq_\lscale{2} \mbox{~\cite{Bali:2012jv}} $& $|V_{ub}|$ & \\ \hline
$-(269(2) \MeV)^3$ & $3.82(20) \times 10^{-3}$  & \\ \hline
\end{tabular}}
\caption{\small Summary of input parameters. Uncertainties in the meson masses are negligible. We set the renormalisation scale $\mu = m_b$, although the scale dependence upon varying $\mu^2 \rightarrow \{0.5,2\} \mu^2$ is negligible since it is $\ORD(\al^2)$ and below numerical precision.}
\label{tab:inputParams}
\end{table}

\subsection{Condensate contributions}
\label{sec:condensates}
\begin{centering}
\begin{figure}[h!]
\begin{overpic}[width=1.0\linewidth]{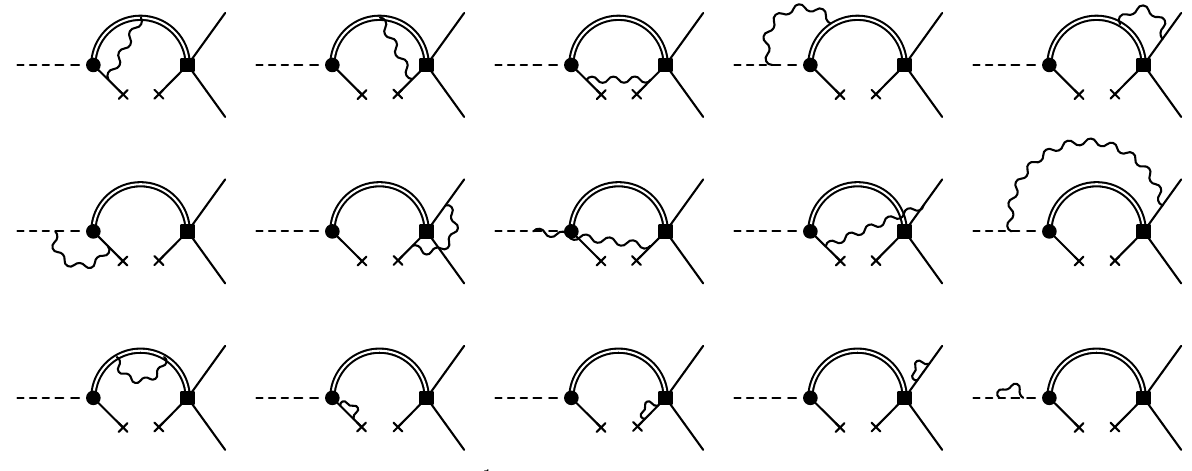}  
\put(10,29){\small $\qbarq$}
\end{overpic}
	\caption{\small  Virtual quark condensate diagrams contributing to $\Pi^{(2)}_{\qbarq}$ in \eqref{eq:mainSR} (the main process). The two crosses denote the quark condensate $\qbarq$.}
	\label{fig:dia-virt-nom-cond}
\end{figure}
\end{centering}

\begin{centering}
\begin{figure}[h!]
\begin{overpic}[width=1.0\linewidth]{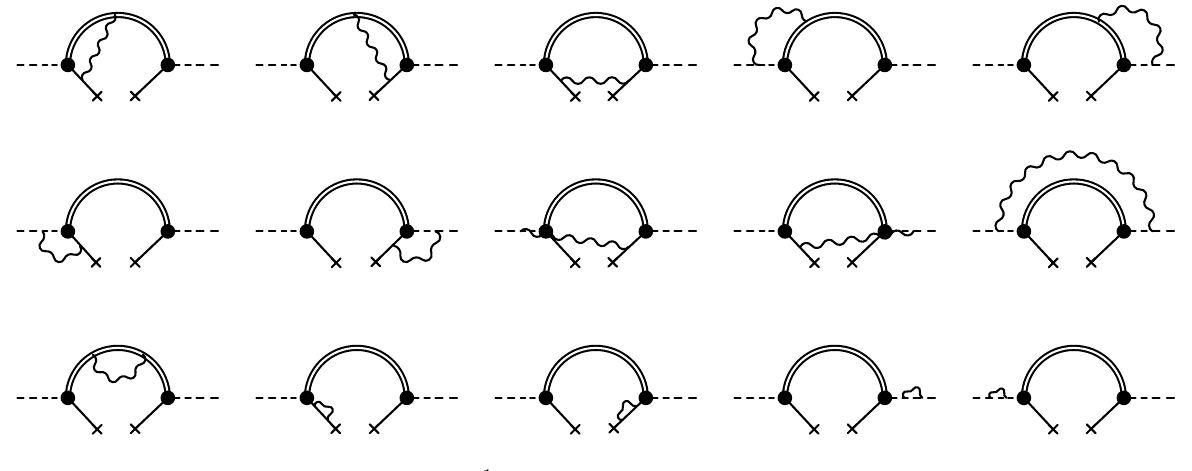} 
\put(8,29){\small $\qbarq$}
\end{overpic} 
	\caption{\small  Quark condensate diagrams contributing to $C^{(2)}_{\qbarq}$ \eqref{eq:C}, 
	that is $|\ZBp|^2$ \eqref{eq:cZB} (the denominator).}
	\label{fig:dia-denom-cond}
\end{figure}
\end{centering}

\begin{centering}
\begin{figure}
\begin{overpic}[width=1.0\linewidth]{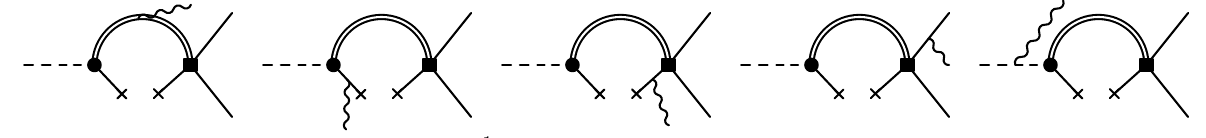}  
\put(9.75,1){\small $\qbarq$}
\end{overpic}
	\caption{\small  Quark condensate diagrams contributing to $\Pi^{\ga}_{\qbarq}$ in \eqref{eq:mainSR} (i.e. the radiative or real emission part). We stress again that the diagrams where the photon connects to the quarks are subsumed by using external form factors \cite{Janowski:2021yvz}.}
	\label{fig:dia-real-nom-cond}
\end{figure}
\end{centering}

This appendix details the contribution of the quark condensate $\qbarq$-diagrams shown in \FIGs\ref{fig:dia-virt-nom-cond}, \ref{fig:dia-denom-cond} and \ref{fig:dia-real-nom-cond}.

On the one hand these diagrams are simpler as they are only one-loop, however the cuts are more involved. Consider the virtual diagrams first.  For the condensates we work in the $m_q= 0$ approximation which 
is well justified as light quark mass effects are small in this case.  
 The LO results are
\begin{align}
    \Im \Pi^{(0)}_{\qbarq} =& - \pi m_+ \geff \qbarq m_\ell \bar u \Gamma v \delta^+(p_B^2-m_b^2) \;, \nonumber \\ 
    \Im C^{(0)}_{\qbarq} =& - \pi m_+^2 m_b \qbarq \delta^+(p_B^2-m_b^2) \;.
\end{align}
Due to the presence of the delta functions, it is often easier to give results after Borel transformation. Gauge invariance of the 1PI diagrams works out similarly to \eqref{eq:NumGI}
\begin{align}
    \Im \Pi^{(2)}_{\qbarq} |_{1-\xi} =& \, \geff m_+ e^2 \qbarq \big( Q_{\ell_1} - Q_b + Q_q \big) \big( Q_\Phi - Q_b + Q_q \big) \frac{m_\ell \bar u \Gamma v}{32 \pi^2 p_B^2} \nonumber \\ & \qquad \qquad  \times \big( B_0 (p_B^2,0,m_b^2) - (m_b^2+p_B^2)C_0(0,p_B^2,p_B^2,0,0,m_b^2) \big)  = 0 \;,
    \end{align}
while for the denominator
\begin{align}
    \Im C^{(2)}_{\qbarq} |_{1-\xi} = & - \frac{m_+^2 e^2}{16\pi^2} m_b \qbarq \big( Q_\Phi - Q_b + Q_q \big)^2 C_0(0,p_B^2,p_B^2,0,0,m_b^2)  =  0 \, .
\end{align}
Here $B_0$ and $C_0$ are the usual Passarino-Veltman functions with \textsc{FeynCalc} normalisation $(2\pi \mu)^{2\epsilon} \int \frac{\dd[d]{k}}{i\pi^2}$. In the condensates in general, and in this $C_0$ in particular, we encounter non-logarithmic (power) singularities at the endpoint $p_B^2 \rightarrow m_b^2$. One can see this by using the derivative trick of \eqref{eq:Ntrick}
\begin{align}
    \Im C_0 ( 0,p_B^2,p_B^2,0,0,m_b^2)  = & \frac{(2\pi \mu)^{2\epsilon}}{i \pi^2} \lim_{N^2 \rightarrow 0} \dv{N^2} \Im \int \frac{\dd[d]{k}}{ ( (p_B-k)^2-m_b^2) ( k^2 - N^2) } \nonumber \\
    = & - \pi \big( 1 + \epsilon ( - \gamma_E + \ln 4 \pi \mu^2 ) \big) \frac{p_B^2+m_b^2}{(p_B^2)^{1-\epsilon}(p_B^2-m_b^2)^{1+2\epsilon}} \; ,
\end{align}
a form seen in \cite{Wang:2015uya}. The divergence as $p_B^2 \rightarrow m_b^2$ is of the IR-type, regulated by $\epsilon <0$. On the other hand the $(p_B^2)^{1-\epsilon}$ part regulates the UV ($\eps>0$) as $p_B^2 \rightarrow \infty$ (after Borel transformation all UV divergences disappear since they are local). 
The IR divergence is manifest after the Borel transform
\begin{equation}
    \mathcal{B}_{M^2} C_0 = e^{\frac{m_B^2-m_b^2}{M^2}} \bigg( \frac{1}{\hat{\epsilon}_\textrm{IR}} + \ln \frac{s_0 \mu^2}{(s_0 - m_b^2)^2} - \int_{m_b^2}^{s_0} \frac{s+m_b^2}{s(s-m_b^2)} \big( e^{\frac{m_b^2-s}{M^2}} - 1 \big) \dd{s} \bigg) \; ,
\end{equation}
where we have used a subtraction method in order to separate out the divergent part of the integral.
 %\footnote{Later we will argue that 
 %$ \frac{1}{\hat{\epsilon}_\textrm{IR}}  \to  \frac{1}{\hat{\epsilon}_\textrm{UV}}$ when considering 
 %$Z$-factors. In the QCD limit they then cancel since $J_B$ does not renormalise.} 
 
The discontinuities are obtained from the standard cutting rules, though the soft slicing takes a different form for the red cuts. As there is one less loop, compared to the perturbative contribution, the red cuts resemble a $1\rightarrow 2$ real decay 
and thus the photon energy is fixed. The $\delta^+(p_B^2-m_b^2) \ln \Delta E_s$ term, generated from the soft region, cancels with a $\theta (p_B^2 - m_b^2 - 2m_b \Delta E_s ) / (p_B^2 - m_b^2)$ only after  Borel transformation, with the $\ln \Delta E_s$ dependence arising  from an incomplete Gamma function. As before, for all diagrams that do not involve a  $q$ quark leg, the soft divergence cancels between red and blue cuts. The denominator $\Phi \, \Phi$- and numerator $\Phi \, \ell_1$-diagrams (which only have red cuts) give physical soft divergences that cancel with the self-energies (denominator) and real radiation (numerator) as in the perturbative calculation. Indeed we recover the universal soft form for the numerator with $\Im \Pi^{(0)}_{\mathbb{1}} \rightarrow \Im \Pi^{(0)}_{\qbarq}$.

The diagrams where the photon connects to a condensate $q$-quark leg are the ones that contribute endpoint IR divergences. Applying charge conservation, the sum of these diagrams is proportional to $Q_q^2$. These IR divergences  cancel 
against $q$-quark self energies $\sim Q_q^2 ( \frac{1}{\epsilon_{\textrm{UV}}} - \frac{1}{\epsilon_{\textrm{IR}}} )$ 
(which formally vanish in DR if $\eps_{\text{UV/IR}}$ is not paid attention to). Now as UV divergences, these terms are removed by the same renormalisations as for the perturbative diagrams serving as a sanity check of the procedure. 

In the numerator a similar slicing procedure can be performed for the collinear logs though now the photon energy is fixed, forcing $z \rightarrow \delta_H = m_b^2 / p_B^2$. There are two $q\, \ell_1$-diagrams (one from each condensate leg) which must be handled with care. 

The real condensate contributions are simpler with $\Im \Pi_{f_B}^{\qbarq} = - \pi \qbarq \delta^+(p_B^2 - m_b^2)$ because all other condensate contributions are contained in the form factors $V_{\para , \perp}$ taken from \cite{Janowski:2021yvz}.

\subsection{More detailed  plots}
\label{app:plots}

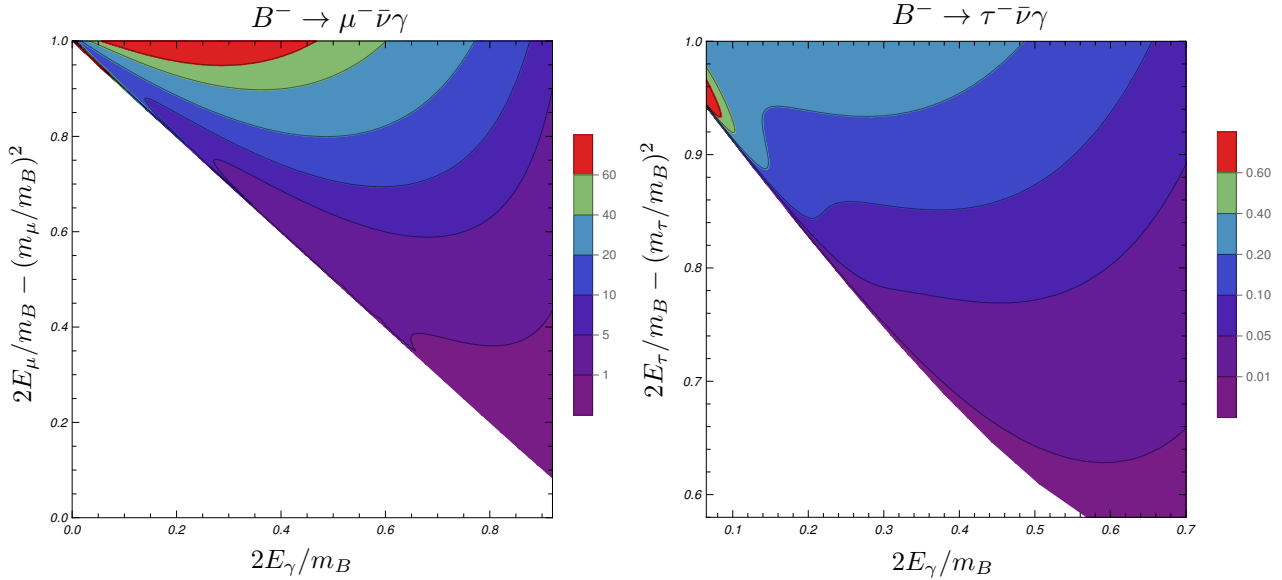
\begin{figure}[t]
\hspace{-1em}
\begin{subfigure}{0.5\linewidth}
\begin{overpic}[width=\linewidth]{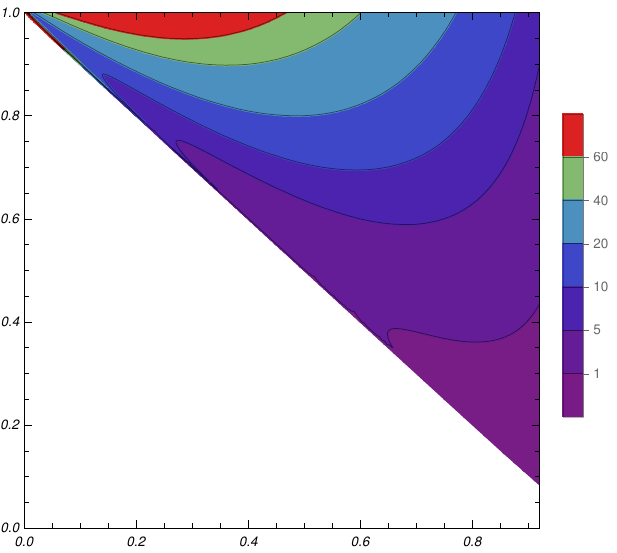} 
%\fboxsep0.4em
%\put(35,91.2){\colorbox{white}{\rule[0em]{40pt}{0em} \, }}
%\put(40,2.8){\colorbox{white}{\rule[0em]{20pt}{0em} \, }}
%\put(-2,35){\colorbox{white}{\rule[0em]{0pt}{1em} \small  \rotatebox{90}{\, \qquad \qquad \qquad  } } }
%\fboxsep0em 
\put(35,90){$B^- \to \mu^- \bar \nu \gamma$}
\put(35,-5){\small $2 E_\gamma / m_B$}
\put(-7,24){\small \rotatebox{90}{$2 E_\mu / m_B - (m_\mu/m_B)^2$ } }
\end{overpic}
\end{subfigure}%
\hspace{2em}%
\begin{subfigure}{0.52\linewidth}
\begin{overpic}[width=\linewidth]{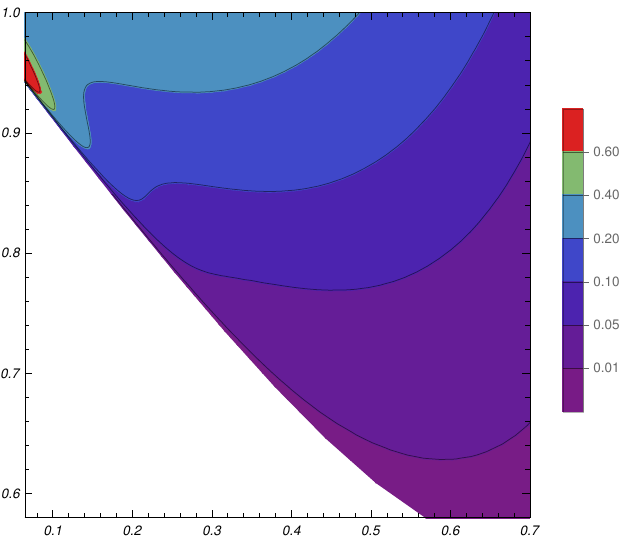}  
%\fboxsep0.4em
%\put(35,89){\colorbox{white}{\rule[0em]{40pt}{0em} \, }}
%\put(40,2.8){\colorbox{white}{\rule[0em]{20pt}{0em} \, }}
%\put(-2,35){\colorbox{white}{\rule[0em]{0pt}{1em} \small  \rotatebox{90}{\, \qquad \qquad \qquad  } } }
%\fboxsep0em
\put(35,87){$B^- \to \tau^- \bar \nu \gamma$}
\put(35,-5){\small $2 E_\gamma / m_B$}
\put(-7,24){\small \rotatebox{90}{$2 E_\tau / m_B - (m_\tau/m_B)^2$ } }
\end{overpic}
\end{subfigure}
\vspace{0.25em}
\caption{\small The Dalitz plot $\frac{1}{\Gamma^{(0)}}\frac{\dd[2]{\Gamma}}{\dd{x}\dd{y}}$ for muons (Left) and taus (Right). The contours show the enhancement from the non-radiative rate. For example on the left the red contour shows where the number of radiative events is $60$ times the number for the non-radiative decay. There are peaks in the soft regions, and where the structure-dependent contributions dominate.}
\label{fig:DalitzPlot}
\end{figure}

\begin{figure}
\hspace{-1em}
\begin{subfigure}{0.5\linewidth}
\begin{overpic}[width=\linewidth]{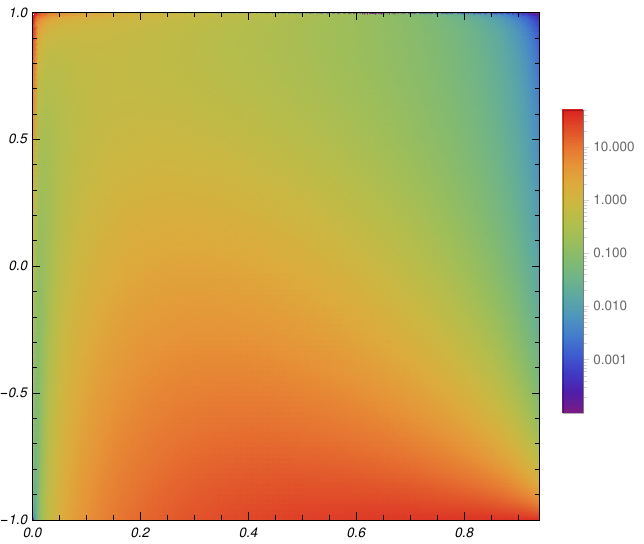} 
%\fboxsep0.4em
%\put(34,84){\colorbox{white}{\rule[0em]{40pt}{0em} \, }}
%\put(40,2){\colorbox{white}{\rule[0em]{20pt}{0em} \, }}
%\put(-2,40){\colorbox{white}{\rule[0em]{0pt}{1em} \small  \rotatebox{90}{\, \qquad } } }
%\fboxsep0em
\put(35,85){$B^- \to \mu^- \bar \nu \gamma$}
\put(35,-5){$ 2 E_\gamma / m_B$}
\put(-5,38){\small \rotatebox{90}{$\cos \theta$ } }
\end{overpic} 
\end{subfigure}%
\hspace{2em}%
\begin{subfigure}{0.512\linewidth}
\begin{overpic}[width=\linewidth]{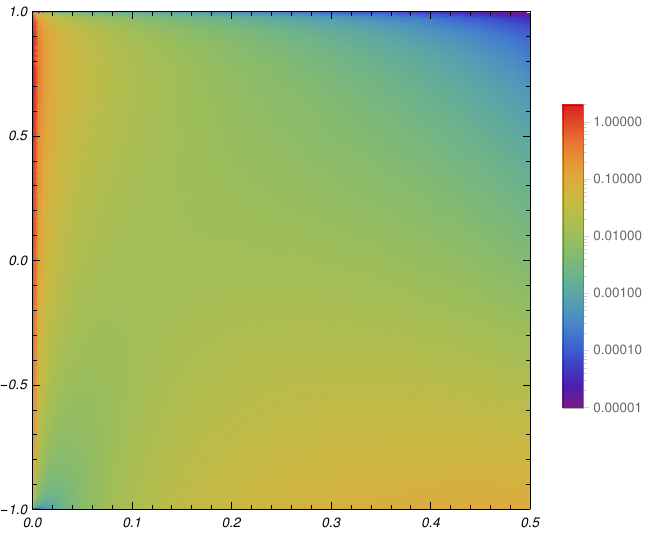}  
%\fboxsep0.4em
%\put(35,85){\colorbox{white}{\rule[0em]{40pt}{0em} \, }}
%\put(40,2){\colorbox{white}{\rule[0em]{20pt}{0em} \, }}
%\put(-2,40){\colorbox{white}{\rule[0em]{0pt}{1em} \small  \rotatebox{90}{\, \qquad } } }
%\fboxsep0em
\put(35,83){$B^- \to \tau^- \bar \nu \gamma$}
\put(35,-5){$2 E_\gamma / m_B$}
\put(-5,38){\small \rotatebox{90}{$\cos \theta $ } }
\end{overpic}
\end{subfigure}
\vspace{0.25em}
\caption{\small The double differential rate $\frac{1}{\Gamma^{(0)}}\frac{\dd[2]{\Gamma}}{\dd{x}\dd{\cos \theta}}$ for muons (Left) and taus (Right). The angle $\theta$ is the angle between the photon and the lepton in the $B$-meson rest frame. Note the colourbar has a different scale for the two channels.}
\vspace{0.25em}
\label{fig:cosPlots}
\end{figure}
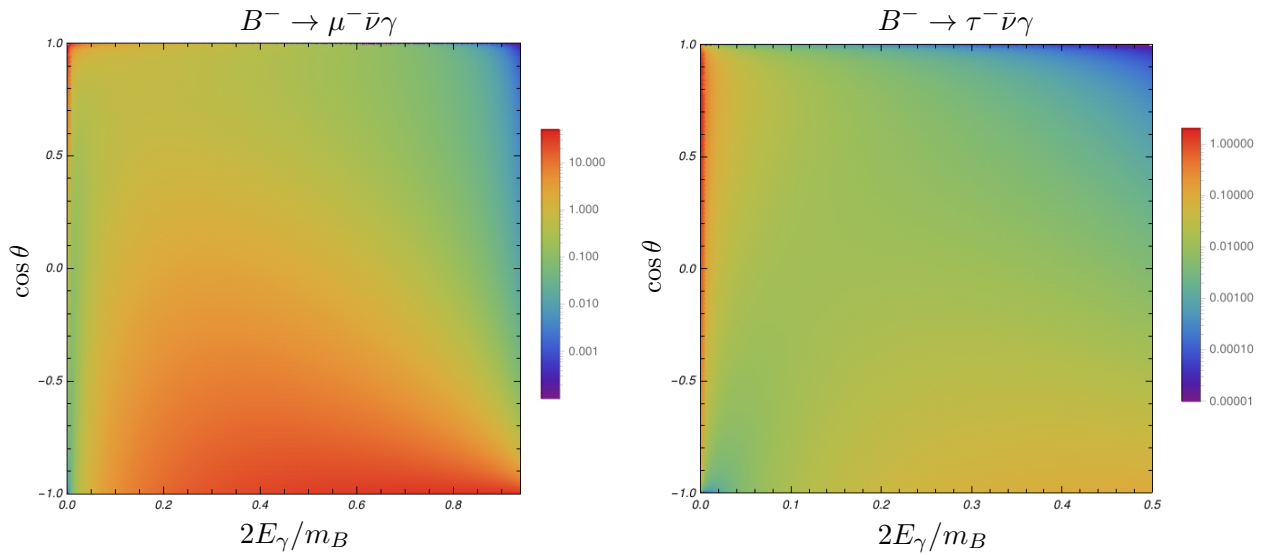

In this appendix, we present  plots with further details. In \FIG\ref{fig:DalitzPlot}, we give the normalised Dalitz plot, that is $\frac{1}{\Gamma^{(0)}}\frac{\dd[2]{\Gamma}}{\dd{x}\dd{y}}$, the double differential decay rate normalised to the LO rate (this type of plot has been shown in \cite{Becirevic:2009aq} for a pole model of form factors). 
Alternatively in \FIG\ref{fig:cosPlots}, the double differential rate is given as $\frac{1}{\Gamma^{(0)}}\frac{\dd[2]{\Gamma}}{\dd{x}\dd{\cos \theta}}$. Here,  the lifting of the helicity-suppression for hard photons is obvious in the muon case. For soft photons, there is a small enhancement visible near $\cos \theta = 1$ which is the collinear region.

In \FIG\ref{fig:noFF} we give the full QED-correction, $\Delta \Gamma_{\text{QED}}$, of \FIG\ref{fig:DeltaPlots} (black line) 
and with real structure-dependent form factors $V_{\perp,\para}$  set to zero (blue line), 
and scalar QED is indicated in red as before.
The difference between the red and blue (dashed) line 
therefore corresponds to the virtual structure-dependent QED-correction (estimated to be $+4.6\%$ and 
$+2.9\%$ for $\mu$ and $\tau$ respectively). 
Conversely, the difference between the black line and the blue line corresponds to the real structure-dependent contribution.\footnote{Note that it is important to use the same scheme, $\overline{\text{MS}}$ in DR in our case, for real and virtual contributions to make the separation well-defined. This applies to scalar QED as there are $1/\eps_{\text{IR}}$-terms and the scheme affects the finite parts, temporarily.}

\begin{centering}
\begin{figure}
\hspace{-1em}
\begin{subfigure}{0.5\linewidth}
\begin{overpic}[width=\linewidth]{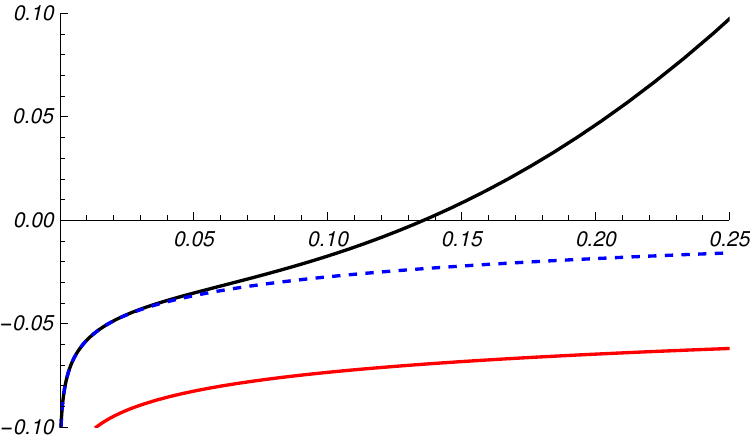}  
\put(33,62){ \large $\boxed{B^- \to \mu^- \bar \nu \gamma}$}
%\fboxsep0.4em
%\put(87.75,25.5){\colorbox{white}{\rule[0em]{15pt}{0em} \, }}
%\put(0,53){\colorbox{white}{\rule[0em]{15pt}{0em} \, }}
\put(83,32){\scriptsize $\decut[\text{GeV}$]}
\put(-2,62){${\Delta \Gamma_{\text{QED}}}$}
%\put(60,5){\small {\color{red} scalar QED}}
%\put(34,35){\scriptsize full QED}
\put(43,4){\scriptsize{\color{red} scalar QED}}
\put(43,18){\scriptsize{\color{blue} no real SD}}
\put(43,44){\scriptsize full QED}
\end{overpic}
\end{subfigure}%
\hspace{1em}%
\begin{subfigure}{0.5\linewidth}
\begin{overpic}[width=\linewidth]{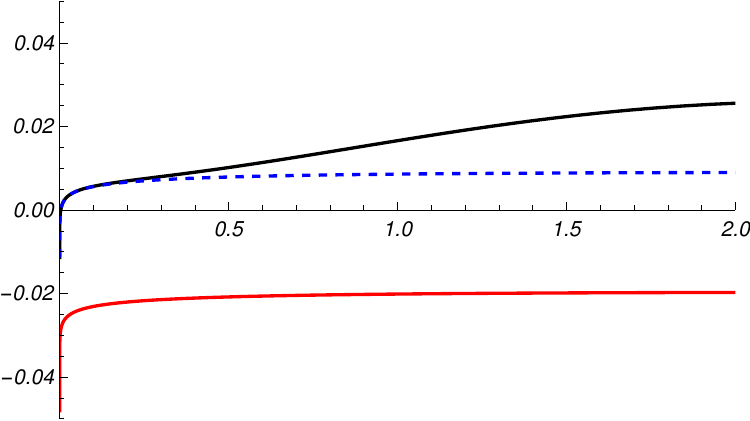}  
\put(33,62){\large $\boxed{B^- \to \tau^- \bar \nu \gamma}$}
%\fboxsep0.4em
%\put(87.5,24){\colorbox{white}{\rule[0em]{20pt}{0em} \, }}
%\put(0,51.5){\colorbox{white}{\rule[0em]{20pt}{0em} \, }}
\put(82,21){\scriptsize $\decut[\text{GeV}]$}
\put(-2,62){  $\Delta \Gamma_{\text{QED}}$}
\put(43,10){\scriptsize{\color{red} scalar QED}}
\put(43,21){\scriptsize{\color{blue} no real SD}}
\put(43,44){\scriptsize full QED}
\end{overpic}
\end{subfigure}
\caption{\small The structure-dependent (SD) contributions to $\Delta \Gamma_{\text{QED}}$, cf. \FIG\ref{fig:DeltaPlots} for muons (Left) and taus (Right). As well as the full result (black) and the scalar QED result (red), we give our calculation having set the real form factors $V_{\para, \perp}$ to zero (blue, dashed). This allows us to gauge the contribution of the real and virtual structure-dependent effects.}
\label{fig:noFF}
\end{figure}
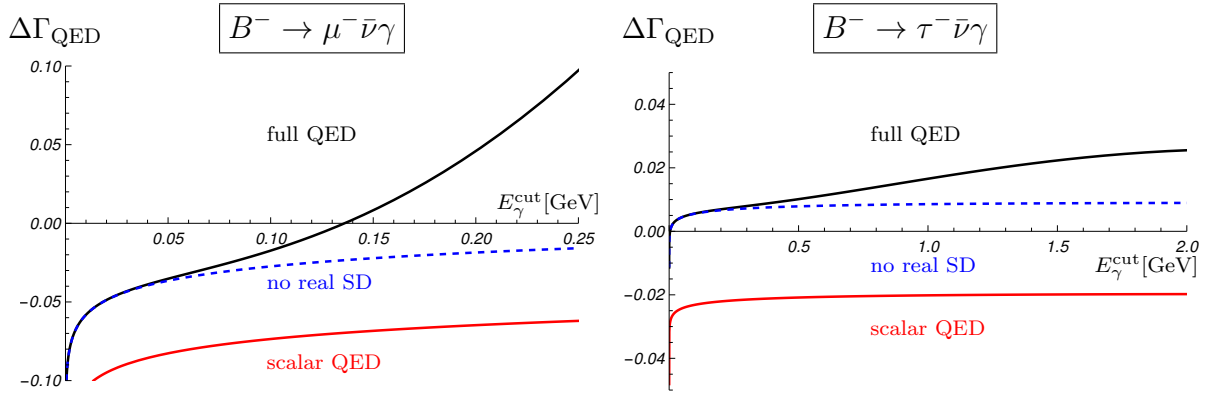
\end{centering}

\section{Further aspects}

\subsection{On the necessity of all cuts}
\label{sec:allCuts}

It is interesting to contrast our approach with the one  
 taken in the calculation of $B \rightarrow \gamma$ form factors $V_{\para, \perp}$ (e.g.  \cite{Janowski:2021yvz}). 
We may consider the situation where we compute the real emission completely within our formalism.\footnote{
This would correspond to  
computing the form factors $V_{\perp,\para}$ explicitly in the soft photon-region. As lamented before, this is not 
easily possible because of the $B^{*,1}$-poles but it is still interesting to reflect on the cuts which is the
aim of this section.}
For the real diagrams, both red (where we cut \emph{after} the photon is radiated) and blue (where we cut \emph{before}), would have to be taken whereas in \cite{Janowski:2021yvz} only the blue cuts are considered. 
How can this be reconciled? 

As discussed in \cite{Nabeebaccus:2022jhu} (Appendix A.1) and in the main text,
the resolution is that the red cuts do not contribute for hard photons ($q^2 \lessapprox 14 \GeV^2$) considered in \cite{Janowski:2021yvz}. To illustrate the differences consider the real amplitude without structure-dependence,  ignoring  lepton radiation, it reads
\begin{multline}
    (\Pi^\gamma)^\rho =   - \gtilde Q_\Phi \frac{p_\Phi^\rho}{p_\Phi \cdot k} \Pi_{f_B} (q^2) m_\ell \, \bar u \Gamma v \, + \\\gtilde Q_\Phi \, \bar u \Gamma_\mu v \bigg( \frac{\Pi_{f_B} (q^2)-\Pi_{f_B} (p_B^2)}{k \cdot p_B} (p_B)^\rho (p_B)^\mu - \Pi_{f_B} (q^2) g^{\rho \mu} \bigg)  + \ldots \, .
\end{multline}
The first term comes from the $\Phi$-diagram with the second line being the contact terms arising from the quark radiation. Taking red cuts is equivalent to cutting $\Pi_{f_B}(q^2)$ while the blue cuts correspond to cutting $\Pi_{f_B} (p_B^2)$. As we take all cuts in our formalism the contact terms are $\mathcal{O}(E_\gamma^0)$ with the $\Phi$-diagram providing the correct Low behaviour as $q^2 \rightarrow p_B^2$. Conversely, if one treats $q^2$ as a separate variable and only cuts in $p_B^2$ (blue cuts only) then $(\Pi_{f_B} (q^2)-\Pi_{f_B} (p_B^2) ) / (k \cdot p_B) \xlongrightarrow{\text{cut}} - \Im \Pi_{f_B} (p_B^2) / (k \cdot p_B)$ which is the analogy of the $\Phi$ term. This resolves the apparent contradiction.
Note that in the $B \rightarrow \gamma$ formalism gauge variant operators are used (no $\Phi$ field) which is acceptable for hard photons as the  $\Phi$-diagram (emission from the $\Phi$) only have red cuts. Our approach is more general as both hard and soft photons can be treated.

On the virtual side taking all the cuts is non-negotiable as it is required for the cancellation of non-physical IR divergences. For example diagrams which connect to the $q$-quark leg generate collinear $\ln m_q$ terms individually in both the red and blue cuts, but these cancel between the cuts
\begin{align}
    \Im \Pi^{(2)}_{\textrm{blue}} |_{\ln m_q} = & + \frac{\alpha}{\pi} Q_q \bigg( Q_{\ell_1} + Q_\Phi - 2 Q_b + \frac{5}{2} Q_q \bigg) \Im \Pi^{(0)}_{m_q \rightarrow 0} \;, \nonumber \\
     \Im \Pi^{(2)}_{\textrm{red}} |_{\ln m_q} = & - \frac{\alpha}{\pi} Q_q \bigg( Q_{\ell_1} + Q_\Phi - 2 Q_b + \frac{5}{2} Q_q \bigg) \Im \Pi^{(0)}_{m_q \rightarrow 0} \;.
\end{align}
As particles are put on-shell when cut, individual cuts also generate soft divergences. For a given diagram in the soft limit, taking the blue and red cuts together, the result is proportional to $\int_k \big( \frac{1}{k^2} + 2\pi i \delta^+(k^2) \big) f$, where $f$ is the rest of the integrand (and diagram-dependent). The $1/k^2$ comes from the blue cut while the delta function arises from the red cut. As $f$ in general contains no poles in the lower half plane, one may evaluate the $k^0$ integral by the residue theorem. The blue cut picks up the only pole at $k^0 = |\bm{k}|-i0$ which puts the photon on-shell and exactly cancels the effect of the $\delta^+(k^2)$. Thus the unphysical soft divergences cancel between blue and red cuts.

\subsection{Toy model}
\label{app:toy}

The methodology of this paper was first tested in a simple toy model 
where the quark loop is replaced by a scalar $B$ propagator as shown in \FIG\ref{fig:redbluecut}. 
Reassuringly, all features such as gauge invariance, the universal soft logs and the universal hard-collinear logs (\SP case) are reproduced.  
It is interesting to look at the hard-collinear logs. 
Specifically, for the virtual 1PI diagrams we find
 \begin{alignat}{2}
\label{eq:toylogs}
& \Pi^{(2), 1\text{PI}}_{\text{Toy,\SP}} |_{\ln m_\ell} &\;=\;& - \frac{\alpha}{\pi} \Pi^{(0)}_{\text{Toy,\SP}} \times \ln m_\ell \;,\nonumber \\[0.1cm]
& \Pi^{(2), 1\text{PI}}_{\text{Toy,\VA}} |_{\ln m_\ell} &\;=\;& \frac{\alpha}{2\pi} \Big( 1 + \mathcal{O}(l_2 \cdot r) \Big) \Pi^{(0)}_{\text{Toy,\VA}} \times \ln m_\ell \;.
\end{alignat}
As expected in the \SP case,  scalar QED  is recovered straightforwardly. To recover  scalar QED  in the \VA case equation \eqref{eq:toylogs} instructs us to take the (sensible) choice $l_2 \cdot r =0$. This extra freedom in the \VA case is a consequence of helicity-suppression and offers a hint at structure-dependent hard-collinear logs.
\begin{table}[h]
	\centering
	\begin{tabular}{ |c||c|c| }
		\hline
		Contribution & S-P & V-A \\
		\hhline{|=||=|=|}
		Real & $-1 +r_E(2-\frac{1}{2}r_E)$ & $-1 +r_E(2-\frac{1}{2}r_E)$ \\
		\hline
		Virtual (1PI) & $-2$ & $1$ \\
		\hline
	    Virtual (Z) & $\frac{3}{2}$ & $\frac{3}{2}$\\
		\hline
	\end{tabular}
	\caption{ \small The table shows the coefficients of $\frac{\alpha}{\pi} \Gamma^0 \ln(m_\ell)$ broken up into the real, virtual (1PI) and virtual (renormalisation) $\mathcal{O}(\alpha)$ contributions in scalar QED. The overall collinear logs are given by summing the columns. In the fully inclusive case 
	($r_E \equiv  2 \decut/ m_B \rightarrow 1$) the \SP $\ln(m_\ell)$ cancel as expected while there are $3$ $m_\ell^2 \log(m_\ell)$ left over in the \VA. Note $- \frac{3}{2} + r_E(2-\frac{1}{2}r_E) $ is just the splitting function in disguise, cf. \eqref{eq:collin}.}
	\label{tab:ScalarQEDLogs}
\end{table}
For the real radiation, discarding the extra structures involving $r$,  scalar QED  is recovered exactly. 
For convenience, an overview of the scalar QED logs is give in \TAB\ref{tab:ScalarQEDLogs} (cf. also  \cite{Zwicky:2021olr}). Note that the \SP and the  \VA real rates are fully identical in scalar QED as a consequence of the equation of motion. This is reflected in the identical real splitting 
function in the table (cf. caption of  \TAB\ref{tab:ScalarQEDLogs}).

\bibliographystyle{utphys}

\bibliography{../../Refs-dropbox/References_QED.bib,../../Refs-dropbox/References_FF-Bgamma.bib}
\end{document}